\documentclass[reprint,superscriptaddress,amsmath,amssymb,aps,showpacs,prb]{revtex4-1}

\usepackage{graphicx}
\usepackage{bm}
\usepackage{epsfig}
\usepackage{amssymb}
\usepackage{amsfonts}
\usepackage{braket}
\usepackage{color}
\usepackage{natbib}
\usepackage[caption=false]{subfig}

\newcommand{\vep}{\varepsilon}

\newcommand{\la}{\langle}
\newcommand{\ra}{\rangle}

\newcommand{\nn}{\nonumber}

\newcommand{\tr}{\textrm}
\graphicspath{{figures/}}

\begin{document}

\date{\today}
\title{RKKY Interactions in Graphene: Dependence on Disorder and Gate
  Voltage}
\author{Hyunyong Lee}
\affiliation{Division of Advanced Materials Science, Pohang University
  of Science and Technology (POSTECH), Pohang 790-784, South Korea}
\author{E. R. Mucciolo}
\affiliation{Department of Physics, University of Central Florida, 
Orlando, Florida 32816, USA}
\author{Georges Bouzerar}
\affiliation{Institut N\'eel, CNRS, d\'epartment MCBT, 25 avenue des
  Martyrs, BP 166, 38042 Grenoble Cedex 09, France}
\affiliation{School of Engineering and Science, Jacobs University
  Bremen, Bremen 28759, Germany} 
\author{S. Kettemann}
\affiliation{Division of Advanced Materials Science, Pohang University
  of Science and Technology (POSTECH), Pohang 790-784, South Korea}
\affiliation{School of Engineering and Science, Jacobs University
  Bremen, Bremen 28759, Germany} 
\email[]{s.kettemann@jacobs-university.de}

\begin{abstract}
  We report the dependence of Ruderman-Kittel-Kasuya-Yoshida\,(RKKY)
  interaction on nonmagmetic disorder and gate voltage in graphene.
  First the semiclassical method is employed to rederive the
  expression for RKKY interaction in clean graphene. Due to the
  pseudogap at Dirac point, the RKKY coupling in undoped graphene is
  found to be proportional to $1/R^3$. Next, we investigate how the
  RKKY interaction depends on nonmagnetic disorder strength and gate
  voltage by studying numerically the Anderson tight-binding model on
  a honeycomb lattice. We observe that the RKKY interaction along the
  armchair direction is more robust to nonmagnetic disorder than in
  other directions. This effect can be explained semiclassically: The
  presence of multiple shortest paths between two lattice sites in the
  armchair directions is found to be responsible for the reduced
  disorder sensitivity. We also present the distribution of the RKKY
  interaction for the zigzag and armchair directions. We identify
  three different shapes of the distributions which are repeated
  periodically along the zigzag direction, while only one kind, and
  more narrow distribution, is observed along the armchair
  direction. Moreover, we find that the distribution of amplitudes of
  the RKKY interaction crosses over from a non-Gaussian shape with
  very long tails to a completely log-normal distribution when
  increasing the nonmagnetic disorder strength. The width of the
  log-normal distribution is found to linearly increase with the
  strength of disorder, in agreement with analytical predictions. At
  finite gate voltage near the Dirac point, Friedel oscillation
  appears in addition to the oscillation from the interference between
  two Dirac points. This results in a beating pattern. We study how
  these beating patterns are effected by the nonmagnetic disorder in
  doped graphene.
\end{abstract}

\maketitle

\section{Introduction}

A large effort has been devoted to the study the electronic transport
properties of graphene due to the unusual nature of the quasiparticles
in this material, which are massless chiral Dirac fermions. 
A recent experiment indicating that vacancies in graphene may induce
local magnetic moments \cite{Chen} renewed the interest in
investigating magnetic properties as well. It is belived that the
carrier-mediated effective exchange coupling between local moments, or
Ruderman-Kittel-Kasuya-Yoshida (RKKY) interaction may play a crucial
role in establishing how magnetism develops in graphene. Probing these
properties locally is not completely out of reach: Using
spin-polarized scanning tunneling spectroscopy, a direct measurement
of the RKKY interaction in a conventional metal has been done by
measuring the magnetization curves of individual magnetic atoms
adsorbed on a platinum surface.\cite{Meier} Similar experiments may
soon be possible on graphene.

Analytical and numerical studies of the RKKY interaction in clean
graphene at the neutrality point have been reported.\cite{Saremi,
  Annica, Sherafati1, Sherafati2} In this context, a dominant feature
of graphene is the bipartite nature of its honeycomb lattice. Due to
particle-hole symmetry of bipartite lattices, the RKKY interaction
always induces ferromagnetic correlation between the magnetic
impurities on the same sublattice, but antiferromagnetic correlation
between the ones on different sublattices.\cite{Saremi} At the
neutrality point, the dependence of the RKKY interaction on the
distance $R$ between two local magnetic moments in graphene is found
to be $1/R^3$, in contrast to the standard behavior in conventional
two-dimensional metallic systems where $1/R^2$ is expected. In doped
graphene, but not too far from the neutrality point, the behavior
changes to $1/R^2$ and two different length scales control the RKKY
interaction: The wavelength corresponding to the inverse of the
distance between the two Dirac points in momentum space, $|\bm{K} -
\bm{K}'|^{-1}$, and the Fermi wavelength $k_F^{-1}$.\cite{Sherafati2}

Since the RKKY interaction is mediated by itinerant electrons in the
host metal, nonmagnetic defects influence directly these
interactions. On-site potential fluctuations scatter the phase of the
electron's wave function as well as cause random changes in its
amplitude, altering any interference phenomenon observed in the clean
system. The effects of weak nonmagnetic disorder on the exchange
interactions in conventional metals have been carefully
studied.\cite{Chatel, Abrahams, Zyuzin, Lerner, Bulaevskii, HYLEE}
Numerical work in the strongly localized regime has been done in
one-dimensional disordered chains.\cite{Sobota} These studies found
that in the diffusive regime the RKKY interaction, when averaged over
disorder configurations, decays exponentially over distances exceeding
the mean free path scale $l_e$. However, its geometrical average value
has the same amplitude as in the clean limit. It is only in the
localized regime, on length scales $R \gg \xi$, i.e., exceeding the
localization length $\xi$, that its geometrical averaged values are
exponentially suppressed so that the RKKY interaction is cutoff over
distances exceeding $\xi$. In our previous work we have studied the
effect of nonmagnetic disorder on the RKKY interaction in undoped
graphene using the kernel polynomial method\,(KPM).\cite{HYLEE} The
unexpected and interesting result found was that, at weak disorder,
the RKKY interaction along the armchair direction is not influenced by
nonmagnetic disorder as much as in the other directions.

Motivated by this unexplained behavior, in this paper we start by
employing the semiclassical method to evaluate the RKKY interaction in
graphene. This calculation helped us understand the $R$-dependence and
the origin of the direction dependence of the sensitivity to disorder
seen in Ref. \onlinecite{HYLEE}. We then used the numerical KPM to
calculate the RKKY interaction in graphene in order to study the
effect of disorder and of doping in more detail.
 
In order to study the disordered conduction electrons in graphene
numerically, we employ the single-band Anderson tight-binding model on
a honeycomb lattice,
\begin{equation}
  \hat{H} = -t\sum_{\la i,j \ra} c_{i}^{\dagger} c_{j} + \sum_{i} (w_i -
  \mu)\,c_{i}^{\dagger} c_{i}, 
  \label{eq:Hamiltonian}
\end{equation}
where $t$ is the hopping integral, $c_i$ is an annihilation operator
of an electron at site\,$i$, $c_i^{\dagger}$ is the corresponding
creation operator, $w_i$ is the uncorrelated on-site disorder
potential distributed uniformly between $[-W/2,W/2]$, and $\la i,j
\ra$ indicates nearest-neighbor sites. Periodic boundary conditions
are used and the lattice constant $a$ and $\hbar$ are set to be unity
in all calculations.

Below, we start with the derivation of the RKKY interaction expression
in clean graphene using the semiclassical method.

\section{Semiclassical approach to the RKKY interaction} 

\begin{figure}[!Ht]
 \captionsetup[subfloat]{font = {bf,up}, position = top} 
\subfloat[~~~~~~~~~~~~~~~~~~~~~~~~~~~~~~] 
{\includegraphics[width=0.22\textwidth]{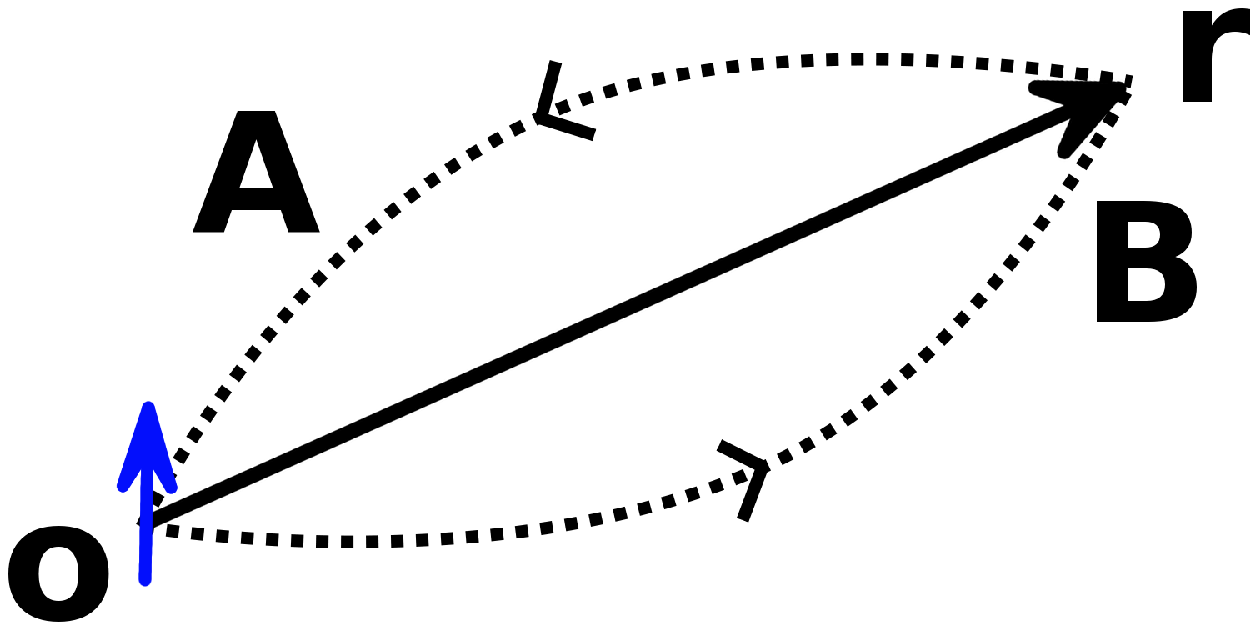}}
\subfloat[~~~~~~~~~~~~~~~~~~~~~~~~~~~~~~]
{\includegraphics[width=0.26\textwidth]{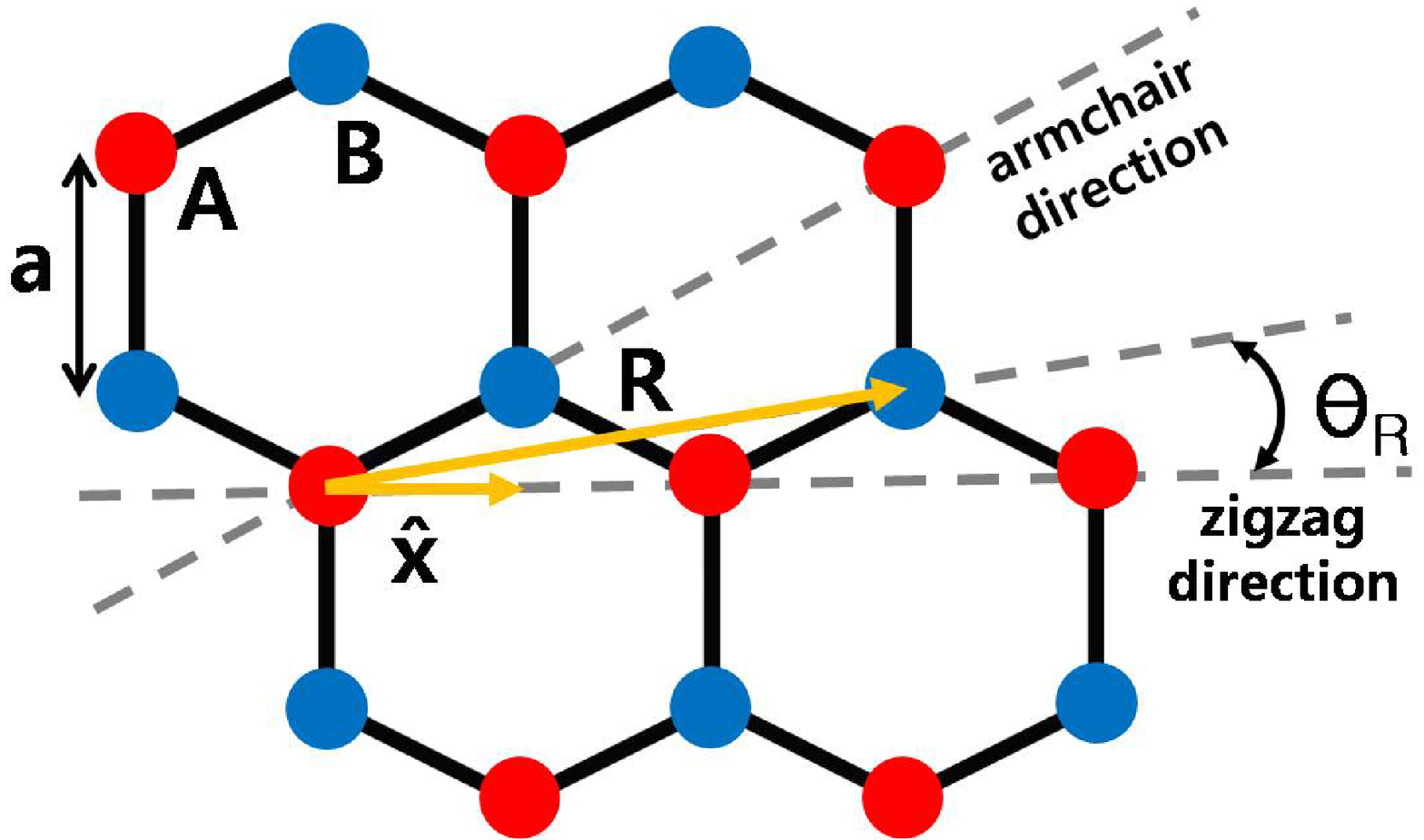}} 
 \caption{(Color online) Schematic diagrams of (a) the propagating
   paths of an electron in clean system and (b) the graphene
   lattice. The two sublattices are denoted as A and B and the two
   representative directions (zigzag and armchair) are indicated as
   dashed gray lines. $\theta_{\bm{R}}$ is the angle between the
   displacement vector of the magnetic impurity $\bm{R}$ and the unit
   vector $\hat{\bm{x}}$.}
\label{fig:schematic1}
\end{figure}
 
Bergmann interpreted the RKKY oscillation as an interference of the
conduction electron's wave function scattered by the magnetic impurity
and calculated the interference using the Huygens' principle in a
three-dimensional metal.\cite{Bergmann} According to the Huygens'
principle in two dimensions,\cite{Cirone} the amplitude of a wave
which propagates from a source at position $\bm{R}'$ decays with
distance and gains a phase factor at a position $\bm{R}$ given by
\begin{equation}
  \Psi(\bm{R}')\,\frac{e^{i \bm{k}\cdot(\bm{R}-\bm{R}')}
  }{\sqrt{|\bm{R} - \bm{R}'|}},
\end{equation}
where $\Psi(\bm{R}')$ is the amplitude of the wave at the source and
the extra factor comes from the asymptotic form of the Bessel function
in two dimensions ($e^{ikr}/\sqrt{r}$). If an electron propagates from
$\bm{R}$ to the origin in graphene (Fig.\,\ref{fig:schematic1}a), the
amplitude gets a phase factor
\begin{equation}
  A = \frac{1}{\sqrt{R}}\,\left[ e^{i(\bm{K}+\bm{q})\cdot{\bm{R}}} +
  e^{i (\bm{K}'+\bm{q}) \cdot{\bm{R}}}  \right] e^{i\vep_q t/2},
\end{equation}
where the wave vector is expanded around the two neighboring Dirac
points $\bm{K}$ and $\bm{K}'$ with a relative wave vector
$\bm{q}$. During its propagation, the electron gets another phase
factor, $e^{i \vep_q t/2}$, where $\vep_q = v_F q$ is the kinetic
energy of the electron near the Dirac point, $v_F$ is the Fermi
velocity and $t/2$ is the propagation time from $\bm{R}$ to the
origin. After the electron is scattered by a magnetic impurity at the
origin, it travels back to the position $\bm{R}$. The amplitude then
gains an additional modulation of the amplitude which is given by
\begin{equation}
  B = \frac{\delta_0}{\sqrt{R}}\,\left[
    e^{-i(\bm{K}+\bm{q})\cdot{\bm{R}}} + e^{-i(\bm{K}'+\bm{q})
      \cdot{\bm{R}}} \right] e^{i(\vep_{q} t/2 + \delta_0)},
\end{equation}
where $\delta_0$ depends on the properties of the magnetic impurity at
the origin. When the electron goes back to $\bm{R}$, its direction is
opposite compared to the first propagation and this is the reason why
the signs are opposite in the phase factors in $A$ and $B$. In the
time domain, however, it propagates in the same direction so that the
phase factor related with the time\,($\vep_q t/2$) has the same sign.
This is consistent with the diagrammatic expression for the RKKY
interaction which has two retarded Green's functions. For a closed
path, a loop of the opposite direction makes a contribution with the
same weight, so that the modulation of the charge density of an
electron which has the energy $\vep_q$ is given by
\begin{eqnarray}
  \Delta\rho_{\vep}(\bm{R}) 
  & = & 2|\psi(\bm{R})|^2 AB \nn \\ 
  & = & \frac{4\delta_0}{V} \frac{ 1+ \cos[(\bm{K}-\bm{K}')\cdot
    \bm{R}]}{R} e^{i \vep_q t},
\end{eqnarray}
such that the total charge modulation is given by
\begin{eqnarray}
 \rho(\bm{R}) & = & \int d\vep\,f(\vep)\,N(\vep) \Delta \rho_{\vep}
 (\bm{R}) \nn \\ & = & - \frac{\delta_0 v_F^2}{V} \frac{ 1+
   \cos[(\bm{K}-\bm{K}')\cdot \bm{R}]}{R^{d+\alpha}},
  \label{eq:charge_density}
\end{eqnarray}
Here, $|\psi(\bm{R})|^2 = 1/V$ for free electron, $V$ is the volume of
the system, $f(\vep)$ is the Fermi-Dirac distribution function,
$N(\vep) = |\vep|^{\alpha}$ is the density of state at the Fermi
level, $\alpha$ is the pseudogap exponent, $d$ is the spatial
dimension, and $t = 2r/v_F$ is the total time it takes to return to
$\bm{R}$.

The total charge density in (Eq.\,\eqref{eq:charge_density}) can be
easily related to the RKKY interaction.\cite{Bergmann} Since in
graphene $d = 2$ and $\alpha = 1$, the resulting distance dependence
is $1/R^3$, which is consistent with previous works.\cite{Saremi,
  Annica, Sherafati1} We conclude therefore that the existence of the
pseudogap at the Dirac point causes the faster spatial decay of the
RKKY interactions in graphene compared to conventional two-dimensional
system ($d=2$, $\alpha = 0$). Even though
Eq.\,\eqref{eq:charge_density} does not give detailed information of
about the RKKY interactions, such as the sign dependence on the
sublattice and an extra phase factor depending on the direction
between point $\bm{R}$ and the origin, it agrees with previous results
obtained by the Green's function method,\cite{Sherafati1,Kogan}
namely,
\begin{eqnarray}
  J_{\textrm{AA}}^{0} & = & -J^2 \frac{1 + \cos[(\bm{K}-\bm{K}')
    \cdot \bm{R}]}{R^3},\\ 
  J_{\textrm{AB}}^{0} & = & J^2 \frac{3 + 3\cos[(\bm{K}-\bm{K}')
    \cdot \bm{R} + \pi -2\theta_R]}{R^3},
  \label{eq:rkky_dirac}
\end{eqnarray}
where $\theta_R$ is an angle between the position vector $\bm{R}$ and
the zigzag-AA direction ($\theta_R = 0)$ as described in
Fig.\,\ref{fig:schematic1}b.

\section{Numerical method}

We start with a general expression for the RKKY exchange coupling
constant in terms of the off-diagonal matrix elements of the local
density of states $\rho_{ij}(E) =
\braket{i|\delta(E-\hat{H})|j}$,\cite{Didier, HYLEE}
\begin{equation}
  J_{\textrm{RKKY}} = -J^2\frac{S(S+1)}{2S^2} \int_{E<\mu} dE
  \int_{E'>\mu} dE' \frac{F(E,E')}{E-E'}, 
  \label{eq:J-KPM}
\end{equation}
where $F(E,E') = \tr{Re}[\rho_{ij}(E)\rho_{ji}(E')]$, $J$ is the local
coupling constant between the localized magnetic impurity and the
itinerant electrons in the host metal, $S$ is the magnitude of the
impurity spin, $i,j$ are the site indices of magnetic impurities
located at position $\bm{R}_i$\,($\bm{R}_j$), and $\mu$ is the Fermi
energy. Zero temperature ($T=0$) is assumed. Using the KPM, one may
calculate the matrix elements of the local density of state
efficiently,\cite{Weisse}
\begin{equation}
  \rho_{ij} \approx \frac{1}{\pi \sqrt{1-E^2}}\left[ g_0\, m^{ij}_0 +
    2 \sum ^{M}_{l=1} g_l\, m^{ij}_l\, T_l(E)\right],
  \label{eq:rho}
\end{equation}
where $T_{l}(E)$ is the $l$th Chebyshev polynomial of the first kind,
$m^{ij}_l = \braket{i|T_l(\hat{H})|j}$, $\hat{H}$ is an electronic
Hamiltonian, and $g_l$ are the Jackson kernels coefficients. The sum
is taken up to a cutoff number $M$. One may obtain the expansion
coefficients $m_{l}^{ij}$ using the recurrence relation of Chebyshev
polynomials, namely, $T_{l+1} = 2 \hat{H} T_l(\hat{H}) -
T_{l-1}(\hat{H})$ with $T_0(\hat{H}) = 1$ and $T_1(\hat{H}) =
\hat{H}$. Implicit in Eq.\,(\ref{eq:rho}) is the normalization of the
energy spectrum to a band of unity width.

\section{Disordered graphene at half filling\,($\mu=0$, $W\neq 0$)}

\begin{figure}[!Ht]
  \captionsetup[subfloat]{font = {bf,up}, position = top,
    captionskip=0pt, farskip=-3pt} 
  \subfloat[~~~~$\theta_R = 0$, zigzag-AA~~~~] 
  {\includegraphics[width=0.24\textwidth]{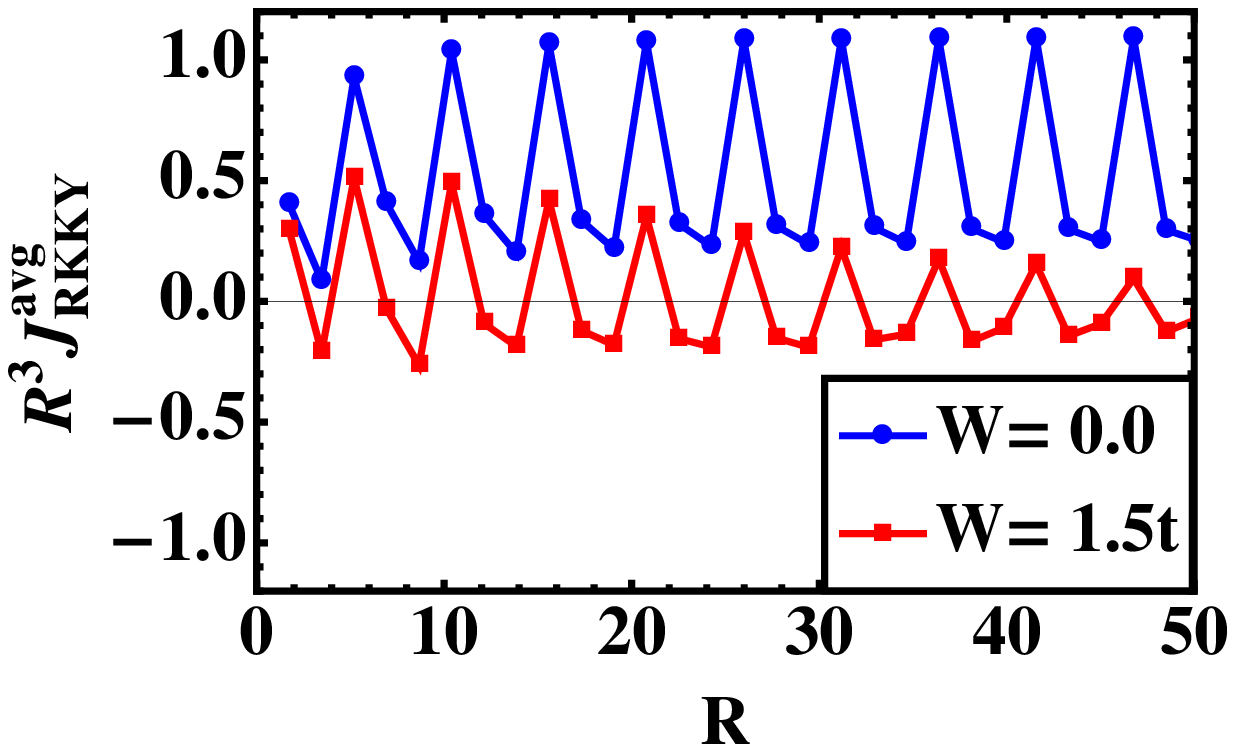}}
  \subfloat[~~~~~~~~~$\theta_R = \pi/12.5$~~~~~~~~~] 
  {\includegraphics[width=0.24\textwidth]{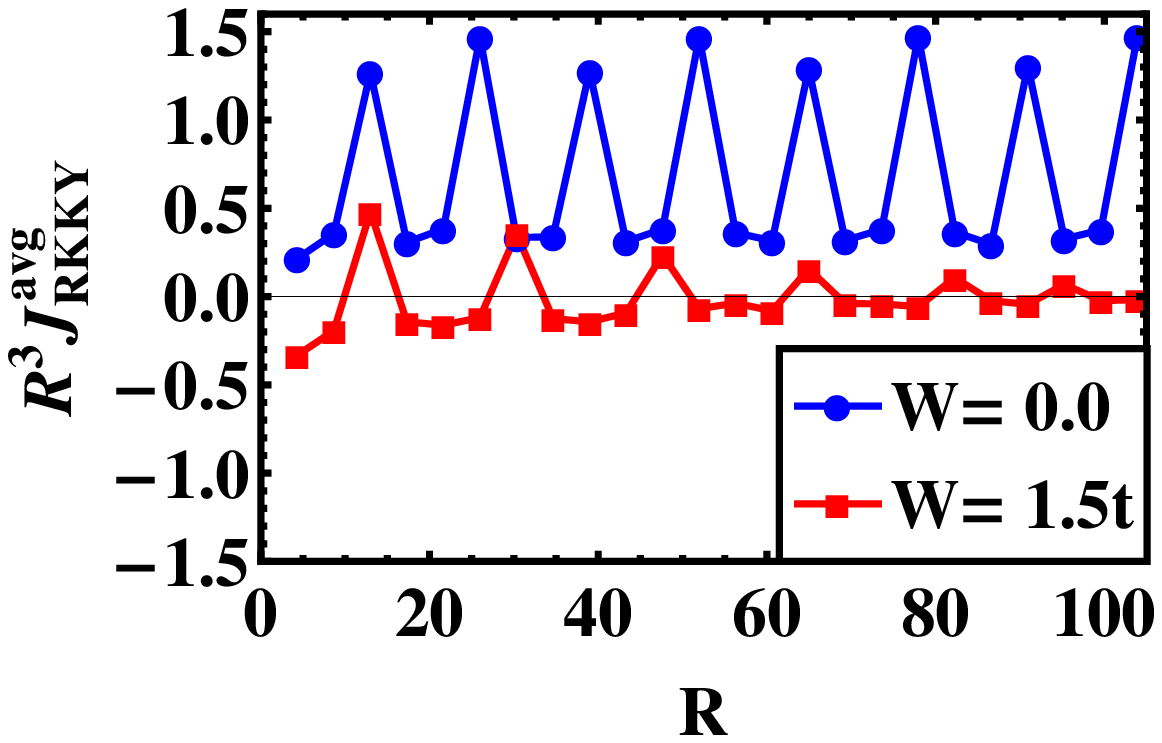}}\\
  \subfloat[~~~~~~~~~$\theta_R = \pi/10$~~~~~~~~~~] 
  {\includegraphics[width=0.24\textwidth]{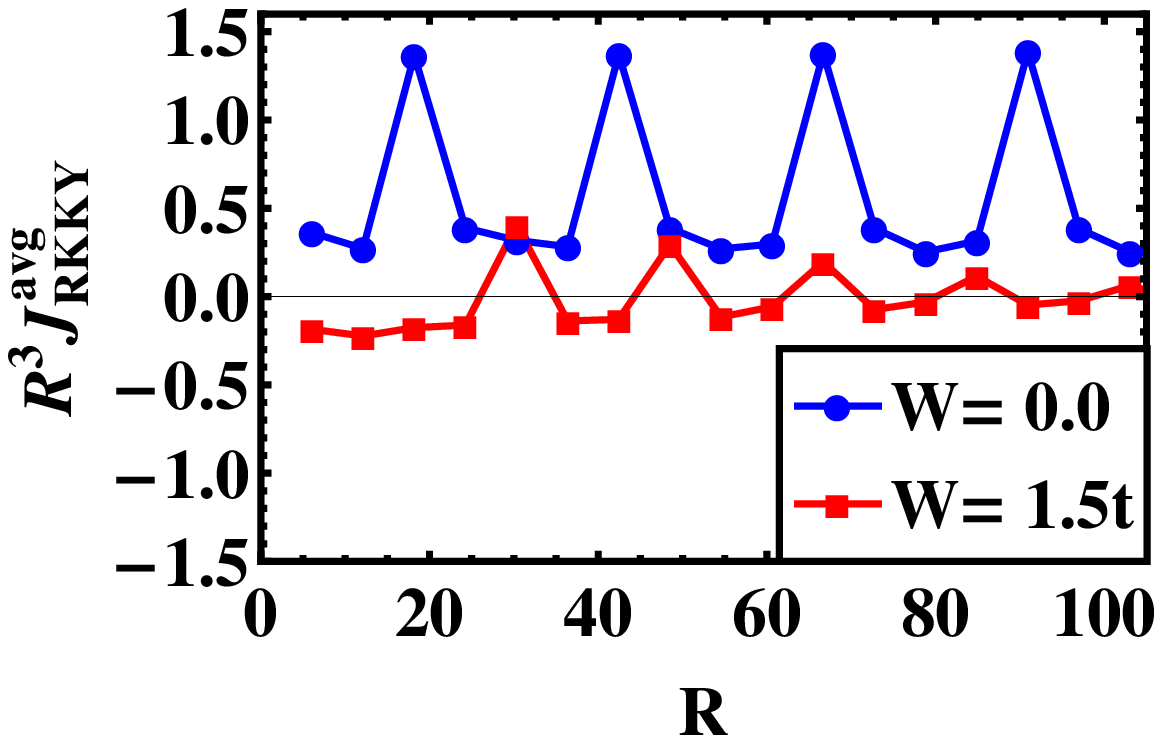}} 
  \subfloat[$\theta_R = \pi/3$, armchair-AA] 
 {\includegraphics[width=0.24\textwidth]{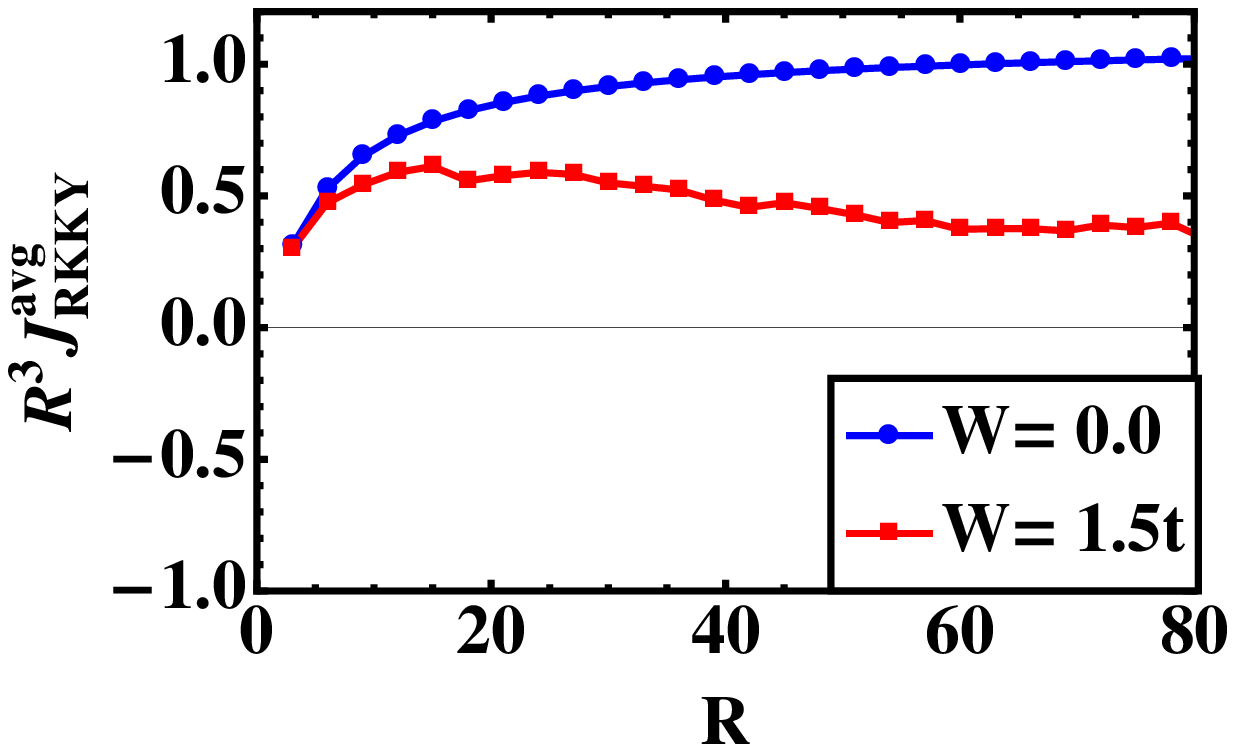}} 
  \caption{(Color online) Plots of the RKKY interaction strength
    multiplied by the cube of the distance, $R^3$, along the (a)
    $\theta_R = 0$ (zigzag-AA) (b) $\theta_R = \pi/12.5$ (c) $\theta_R
    = \pi/10$ and (d) $\theta_R = \pi/3$ (armchair-AA) direction in
    the diffusive regime, as averaged over 1600 different disorder
    configurations. A lattice with $5\times 10^5$ sites and a
    polynomial degree cutoff of $M = 5\times 10^3$ are used in these
    numerical calculations.}
  \label{fig:avg_angle} 
\end{figure}

We have investigated the effect of on-site nonmagnetic disorder at the
charge neutrality point\,($\mu=0,\,W\neq 0$). For each value of $W$,
1600 different disorder configurations were generated and the
corresponding matrix elements of the density of state $\rho_{ij}$
(Eq.\,\ref{eq:rho}) evaluated through the KPM with a Chebyshev
polynomial cutoff number $M=5\times 10^3$ on a lattice with $5\times
10^5$ sites. We have previously reported that in the diffusive
regime\,($l_e<R<\xi$), which was characterized by determining the mean
free path $l_e$ and the localization length $\xi$, the armchair
direction is the only direction in which the averaged value of the
RKKY interaction over disorder configurations does not alter its
sign.\cite{HYLEE} To illustrate this effect, we calculated the RKKY
interactions along the directions ($\theta_R = \pi/12.5,\, \pi/10$)
which pass through only the same sublattice and the results are shown
in Fig.\,\ref{fig:avg_angle} together with the ones for the zigzag-AA
and armchair-AA directions. Notice that the amplitudes of the RKKY
interaction are multiplied by the cube of the distance in order to make
the effect of nonmagnetic disorder more transparent. As expected from
previous studies,\cite{Chatel, Abrahams, Zyuzin, Lerner, Bulaevskii}
the amplitude of the averaged interactions decays exponentially over
length scales exceeding the mean free path $l_e$,
\begin{equation}
\la J_{\tr{RKKY}} \ra_{\tr{avg}} = J_{\tr{RKKY}}^{\tr{clean}}\, e^{-R/l_e},
\end{equation}
where $\la O \ra_{\tr{avg}}$ indicates averaging over disorder
configurations and $J_{\tr{RKKY}}^{\tr{clean}}$ is the amplitude of
the RKKY interaction in a clean system. On-site disorder breaks the
sublattice symmetry. Consequently, the sign of the RKKY interaction
between the moments which are localized at the same sublattice
fluctuates, allowing both ferromagnetic and antiferromagnetic
correlation (Fig.\,\ref{fig:avg_angle}a,\,b,\,c). The importance of
the sublattice symmetry was highlighted in our previous work by
considering the hopping disorder.\cite{HYLEE} When randomness is added
to the hopping integral $t = t_0 + \Delta t$ without on-site disorder,
it does not break the sublattice symmetry and the RKKY interaction
never changes its sign for magnetic moments sitting in either the same
or different sublattices, even for a fixed disorder
configuration. Interestingly, the averaged RKKY interaction amplitude
along the armchair direction with on-site disorder does not change
sign (Fig.\,\ref{fig:avg_angle}d), while for a particular disorder
configuration it randomly changes both sign and amplitude.

\begin{figure}[!Ht]
 \captionsetup[subfloat]{font = {bf,up}, position = top,
   captionskip=0pt, farskip=0pt} 
 \subfloat[zigzag~~~~~]
 {\includegraphics[scale=0.65]{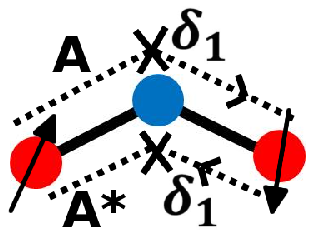}} 
 \subfloat[armchair~~~~~~] 
 {\includegraphics[scale=0.6]{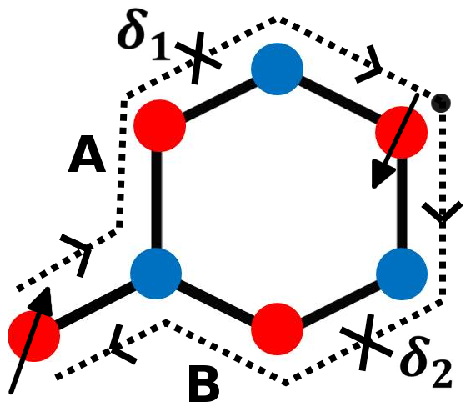}}
 \caption{(Color online) Schematic diagrams of the shortest path from
   the origin to the first lattice point along (a) the zigzag-AA
   direction and (b) the armchair-AA direction. $\delta_{1\,(2)}$
   denotes the phase shift due to nonmagnetic disorder and ${\bf
     A\,(B)}$ denote the modulation of the charge density amplitude.}
 \label{fig:schematic2}
\end{figure}

In a dirty metal the charge density is modified by disorder due to the
random phase factors. When these random phase factors are simply
averaged over disorder configurations, they gain an exponentially
decaying factor. Due to the geometrical anisotropy in graphene, the
influence of these random phase factors depends on the path
direction. According to Feynman's path integral representation, the
coupling is dominated by the shortest path between two magnetic
impurities. There always exists an even number of shortest paths along
the armchair direction (Fig.\,\ref{fig:schematic2}b), but only one
along the zigzag direction (Fig.\,\ref{fig:schematic2}a). Therefore,
for the zigzag direction, the electron wave is scattered by the same
disorder twice with the same phase (e.g. $\delta_1$ in
Fig.\,\ref{fig:schematic2}a) when it returns to the origin. However,
along the armchair direction, there are closed paths which include
different impurities and thereby different scattering phases
(e.g. $\delta_1$ and $\delta_2$ in Fig.\,\ref{fig:schematic2}b). If we
average the modulation of the charge density over the phases
$\delta_1$ and $\delta_2$ we find
\begin{equation}
  \la \Delta \rho_{\tr{zg}}(\bm{R})\ra = \Delta \rho_{0}\, \la
  e^{i2\delta_1} \ra = \Delta \rho_0\, e^{-2\la \delta_1^2 \ra}
  \label{eq:mod_zg}\\
\end{equation}
and
\begin{eqnarray}
  \la \Delta \rho_{\tr{arm}}(\bm{R})\ra &=& \Delta \rho_{0}\, \la
  e^{i2\delta_1} + e^{i(\delta_1+\delta_2)} +e^{i2\delta_2}\ra
  \nn\\ &=& \Delta \rho_{0}\, [\, 2\,e^{-2\la \delta_1^2 \ra} +
    e^{-\la \delta_1^2 \ra}\,],
  \label{eq:mod_arm}
\end{eqnarray}
where $\la \cdots \ra$ is the average over disorder configurations and
$\Delta\rho_0$ is the modulation of the charge density in the clean
system [Eq.\,\eqref{eq:charge_density}]. The modulation along the
armchair direction [Eq.\,\eqref{eq:mod_arm}] is dominated by the term
$\Delta\rho_0\,e^{-\la \delta_1^2 \ra}$ coming from the closed loop
connected by two unrelated paths, so that the average value is
exponentially closer to the clean value, while the zigzag direction is
dominated by $\Delta\rho_0\,e^{-2\la \delta_1^2 \ra}$
[Eq.\,\eqref{eq:mod_zg}]. In this sense, one may expect the RKKY
interaction in the armchair direction to be less sensitive to the
nonmagnetic weak disorder than in other directions. This may explain
our previous results \cite{HYLEE} and the results shown in
Fig.\,\ref{fig:avg_angle}, where the averaged value in the armchair
direction remains susbstantially larger than in the other directions
and does not change its sign.

\begin{figure}[!Ht]
  \captionsetup[subfloat]{font = {bf,up}, position = top,
    captionskip=0pt, farskip=0pt} 
  \subfloat[~~~~~~~~~~zigzag1~~~~~~~~~~~~] 
  {\includegraphics[width=0.24\textwidth]{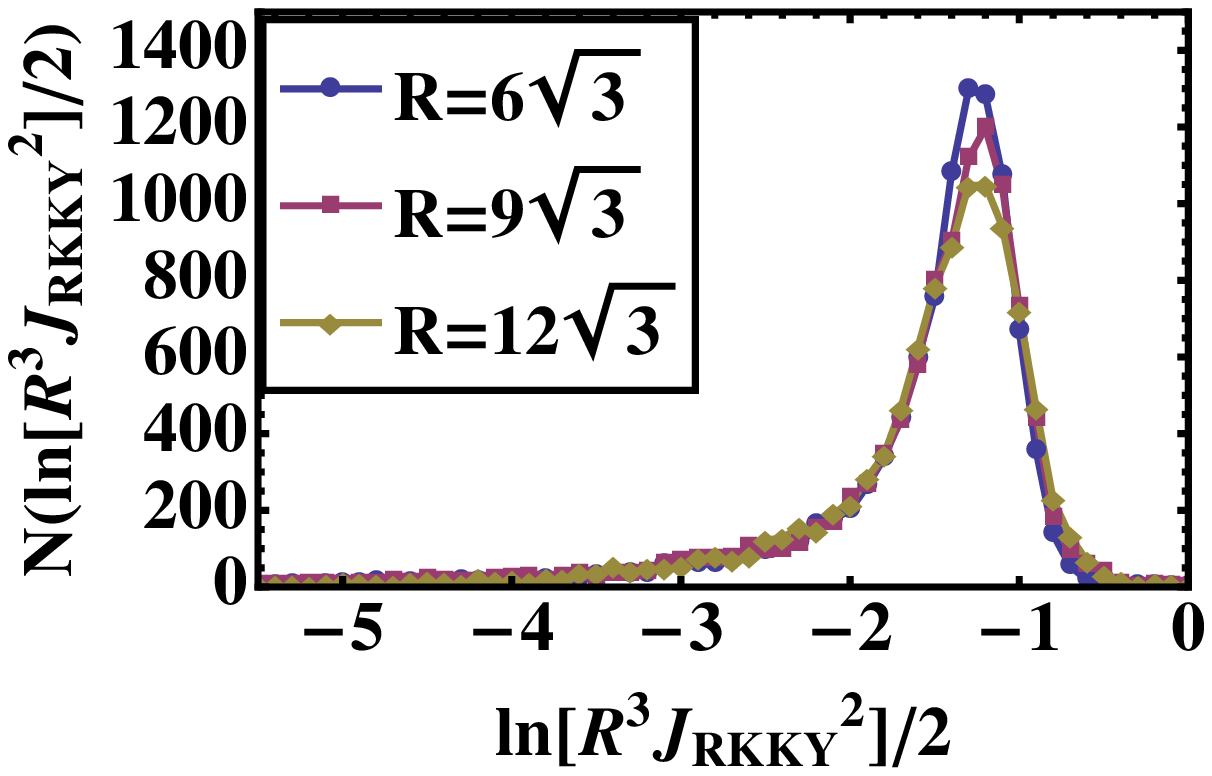}}
  \subfloat[~~~~~~~~~~zigzag2~~~~~~~~~~~~] 
  {\includegraphics[width=0.24\textwidth]{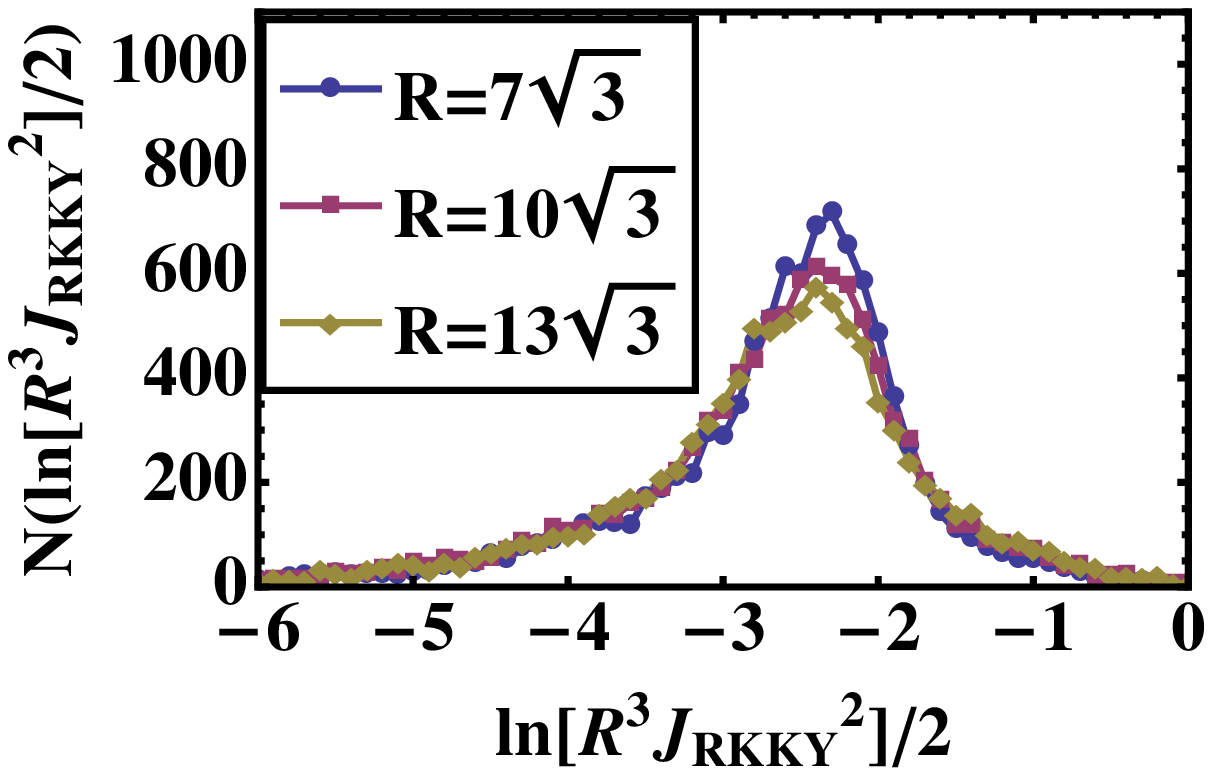}}\\
  \subfloat[~~~~~~~~~~zigzag3~~~~~~~~~~~~] 
  {\includegraphics[width=0.24\textwidth]{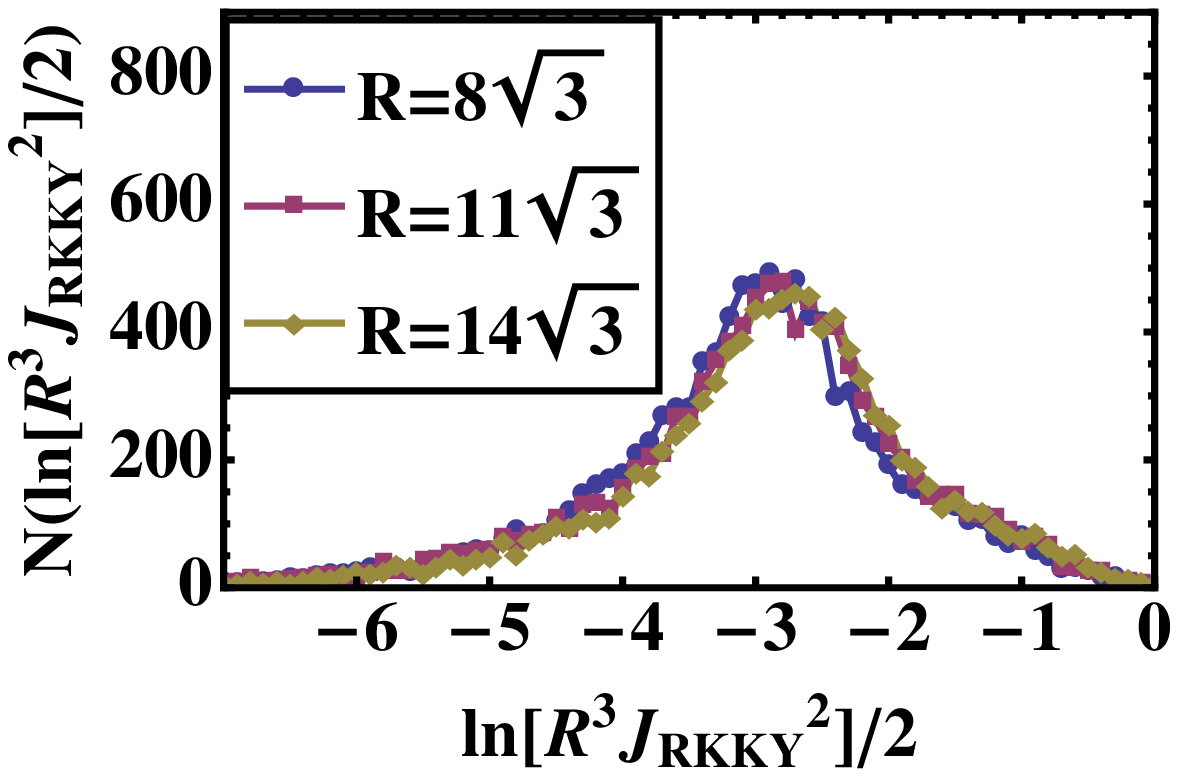}} 
  \subfloat[~~~~~~~~~armchair~~~~~~~~~~~~] 
  {\includegraphics[width=0.24\textwidth]{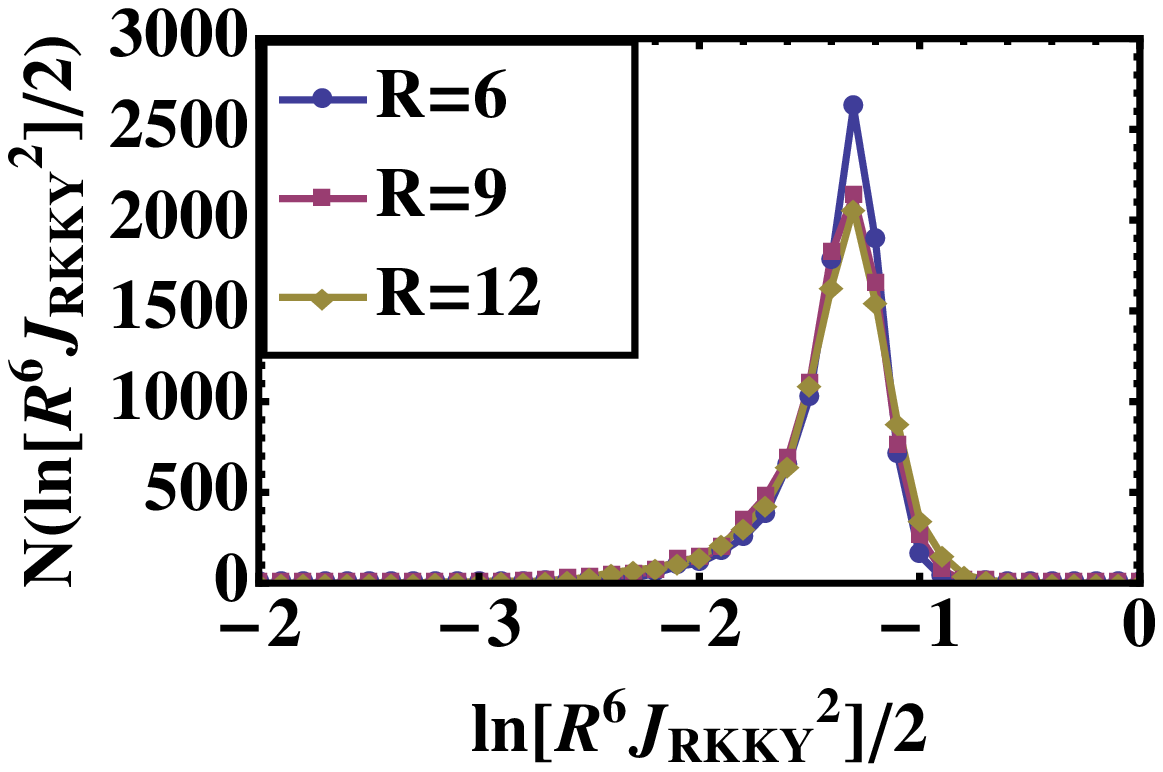}}\\
  \subfloat[~~~~~~~~~~~~~~~~~~~~~~~~~~~~~~~~] 
  {\includegraphics[width=0.24\textwidth]{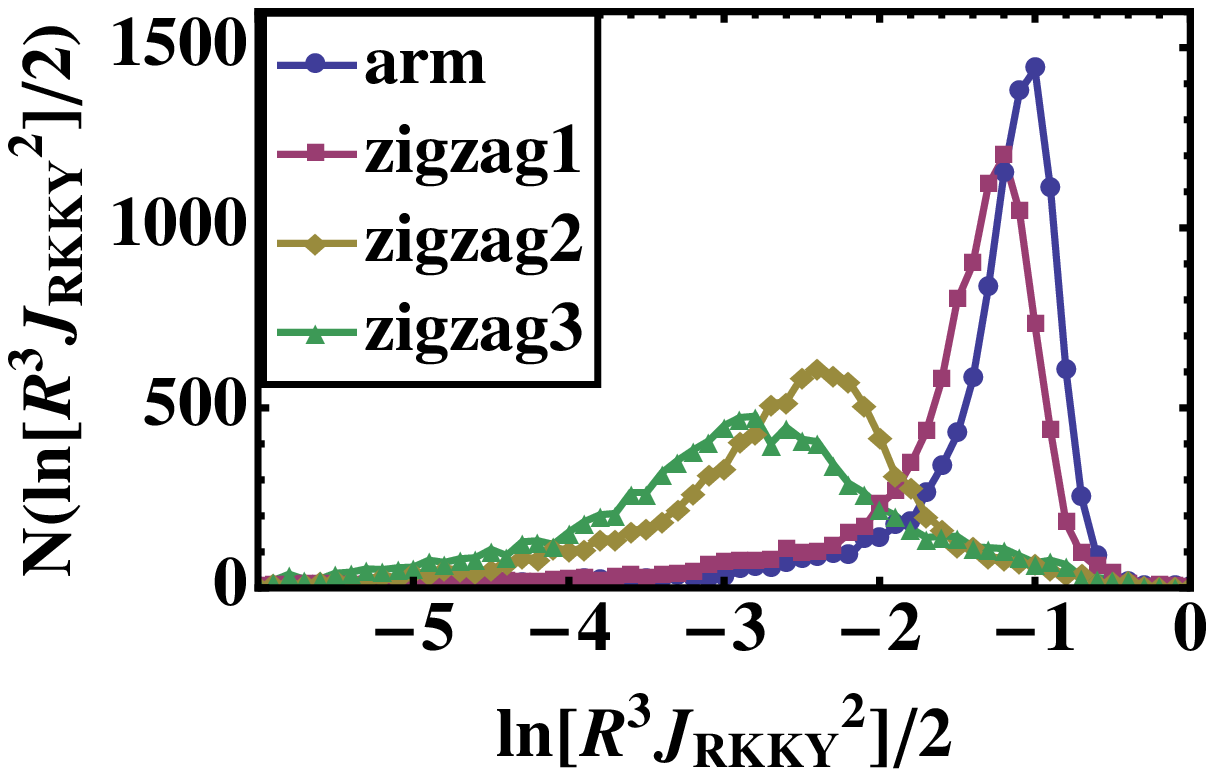}}
  \subfloat[~~~~~~~~~~~~~~~~~~~~~~~~~~~~~~] 
  {\includegraphics[width=0.22\textwidth]{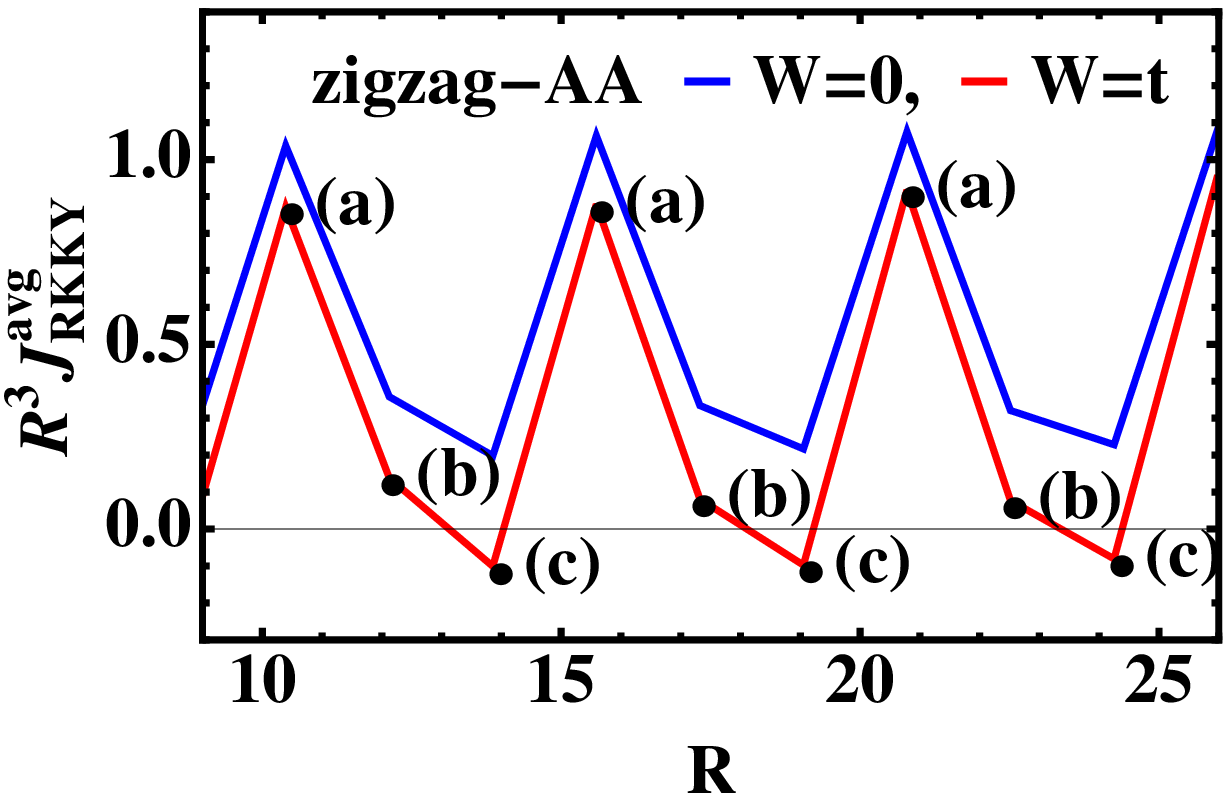}}
  \caption{(Color online) Plots of the distribution of the RKKY
    interaction amplitude\,($\sqrt{J_{\tr{RKKY}}^2}$) multiplied by
    the cube of distance, $R^3$, for the (a) first (b) second (c)
    third of the triplet of points along zigzag-AA direction\,(see
    Fig. 4 f for the definition of these points), and (d) for the
    armchair direction. The disorder strength is fixed to $W = t$.
    For comparison, these distributions are plotted together in (e). A
    lattice with $2\times 10^4$ sites and a polynomial degree cutoff
    of $M=10^3$ are used. The lattice constant $a$ is set to unity.}
  \label{fig:distributions_weak} 
\end{figure}

In order to investigate in more detail how much nonmagnetic disorder
affects the RKKY interaction and how the latter depends on direction
in the graphene lattice and on particular lattice points, we have
calculated the distribution of the interaction amplitude for a
disorder strength $W = t$ and at lattice points which are on the
zigzag or armchair direction. The results are shown in
Fig.\,\ref{fig:distributions_weak} a-e. In these calculations,
$3\times 10^4$ different disorder configurations are used with for a
lattice of $2\times 10^4$ sites, while the number of Chebyshev
polynomials $M=10^3$ is fixed. When the distance $R$ between the
magnetic impurities is smaller than the mean free path, the RKKY
interaction multiplied by the cube of the distance has a shape which
does not depend on the distance. Interestingly, the distribution has
three different shapes along the zigzag-AA direction
(Fig.\,\ref{fig:distributions_weak}a-c), which are repeated
periodically every third site in that direction. However, there is
only one type of distribution along the armchair direction, as can be
seen in \,(Fig.\,\ref{fig:distributions_weak}d). The oscillating
factor in the semiclassical expressions for the charge density,
$\cos[(\bm{K}-\bm{K}')\cdot \bm{R}]$ in Eq. (\ref{eq:charge_density}),
takes only the values ,$(1,-1/2,-1/2)$ along sites in the zigzag-AA
direction. The RKKY interaction on sites which give the same numerical
factor, either $1$ or $-1/2$, are found to have the same
distribution. In order to directly compare them with each other, we
plot together the distributions of the RKKY interactions obtained for
these sites in Fig.\,\ref{fig:distributions_weak}e. The RKKY
interaction at sites where the oscillating factor takes the value
(-1/2) along the zigzag-AA direction has the broadest\,(green)
distribution, while the RKKY interaction at site where the oscillating
factor takes the value (1) has the narrowest\,(red) distribution.

The unaveraged RKKY interaction, i.e., that obtained for a particular
disorder representation, is found to remain long ranged as in the
clean system. However, we find that the oscillations acquire a random
phase that also changes with the distance $R$ between magnetic
impurities. Therefore, the average value does not characterize the
RKKY interaction at distances exceeding the mean free path. To find
the typical value, we evaluated the geometrical average of the
interaction amplitude, which is defined by

\begin{equation}
  J_{\tr{RKKY}}^{\tr{geo}} \equiv \exp\Big[\frac{1}{2} \la \,
  \ln(J_{\tr{RKKY}})^2\, \ra_{\tr{avg}} \Big].
\end{equation}

\begin{figure}[!Ht]
  \captionsetup[subfloat]{font = {bf,up}, position = top,
    captionskip=0pt, farskip=-5pt} 
  \subfloat[~~~~~~~~~~~~~~~~~~~~~~~~~~~~~~~~~~~~~~~~~~~~~~~~~~~~] 
  {\includegraphics[width=0.35\textwidth]{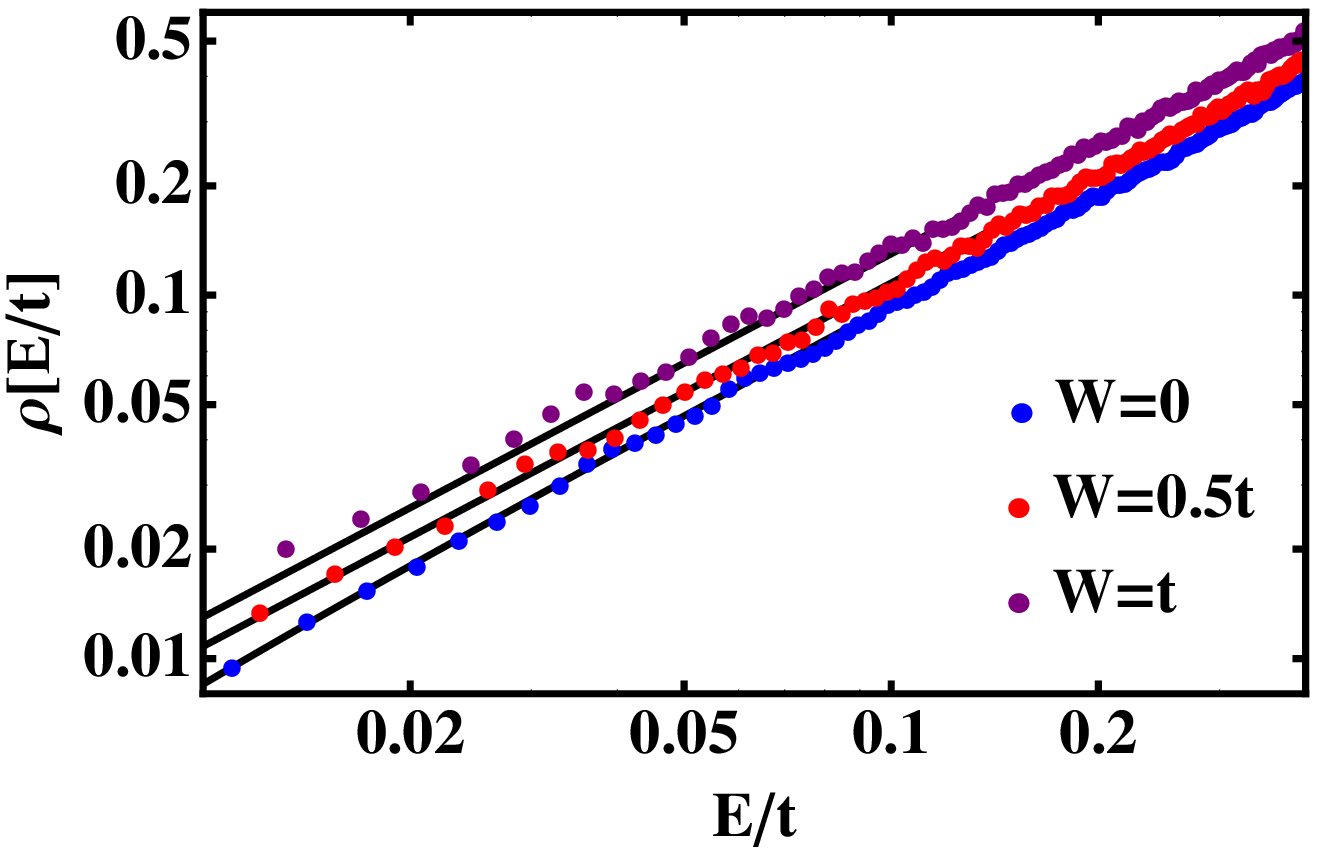}} \\
  \subfloat[~~~~~~~~~~~~~~~~~~~~~~~~~~~~~~~~~~~~~~~~~~~~~~~~~~~~] 
  {\includegraphics[width=0.4\textwidth]{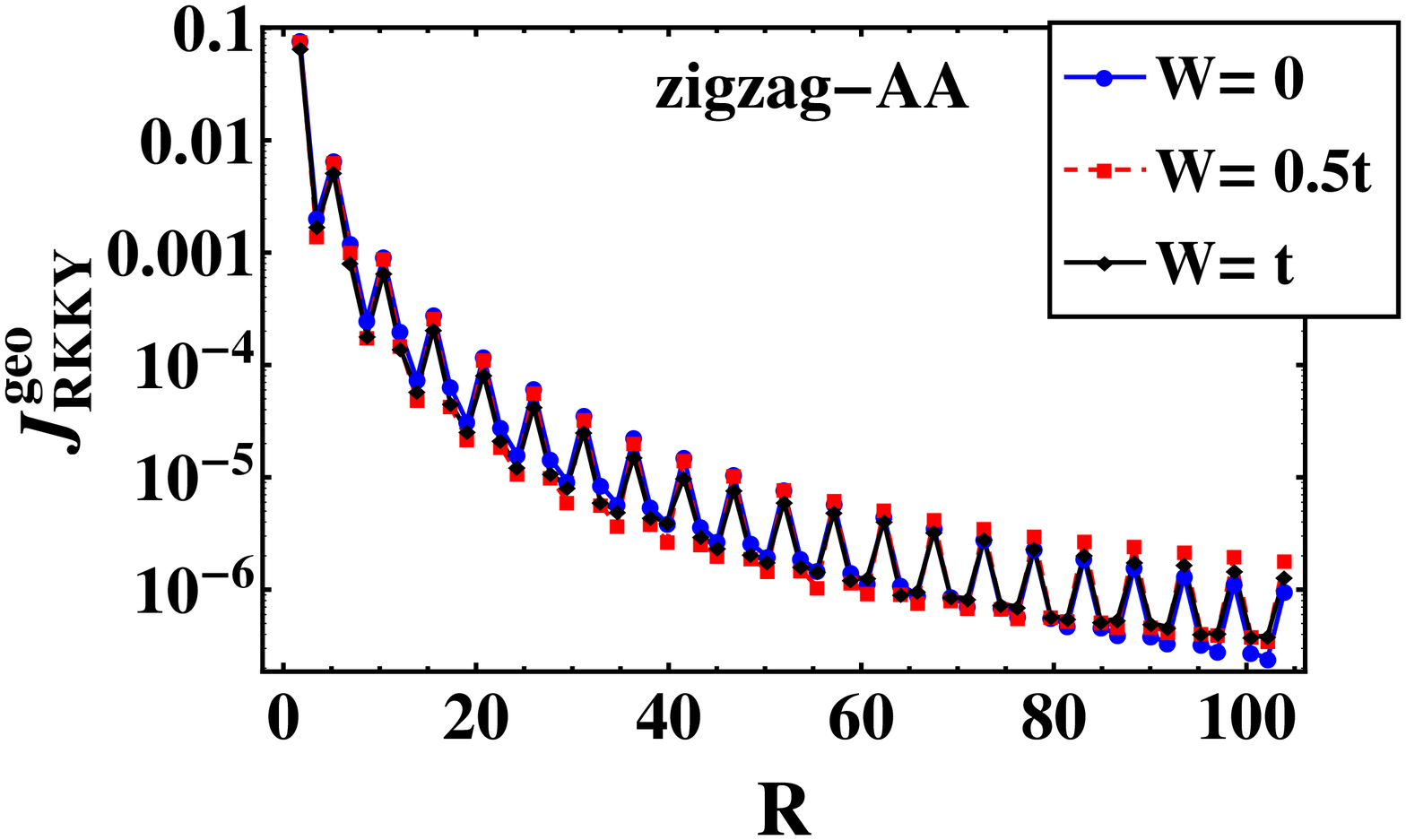}}
  \caption{(Color online) (a) The averaged density of states in a
    double logarithmic plot (b) and the geometrical averaged RKKY
    interactions along the zigzag-AA direction. For the density of
    state calculation, a lattice with $3\times 10^7$ sites and a
    polynomial degree cutoff of $M = 7\times 10^3$ are used. For the
    RKKY interaction calculation, the same lattice size and polynomial
    cutoff of Fig.\,\ref{fig:avg_angle} are used. The black lines in
    (a) represent fittings to the relation $\rho(E) = \gamma |E|$ with
    $\gamma = 0.95,\,1.07,\,1.3$ for $W = 0,\,0.5t, t$, respectively.}
  \label{fig:geo_avg} 
\end{figure}
 
\begin{figure}[!Ht]
 \captionsetup[subfloat]{font = {bf,up}, position = top,
   captionskip=0pt, farskip=-5pt} 
 \subfloat[~~~~~~~~~~~~~~~~~~~~~~~~~~~~~~~~~~~~~~~~~~~~~~~~~~~~] 
 {\includegraphics[width=0.4\textwidth]{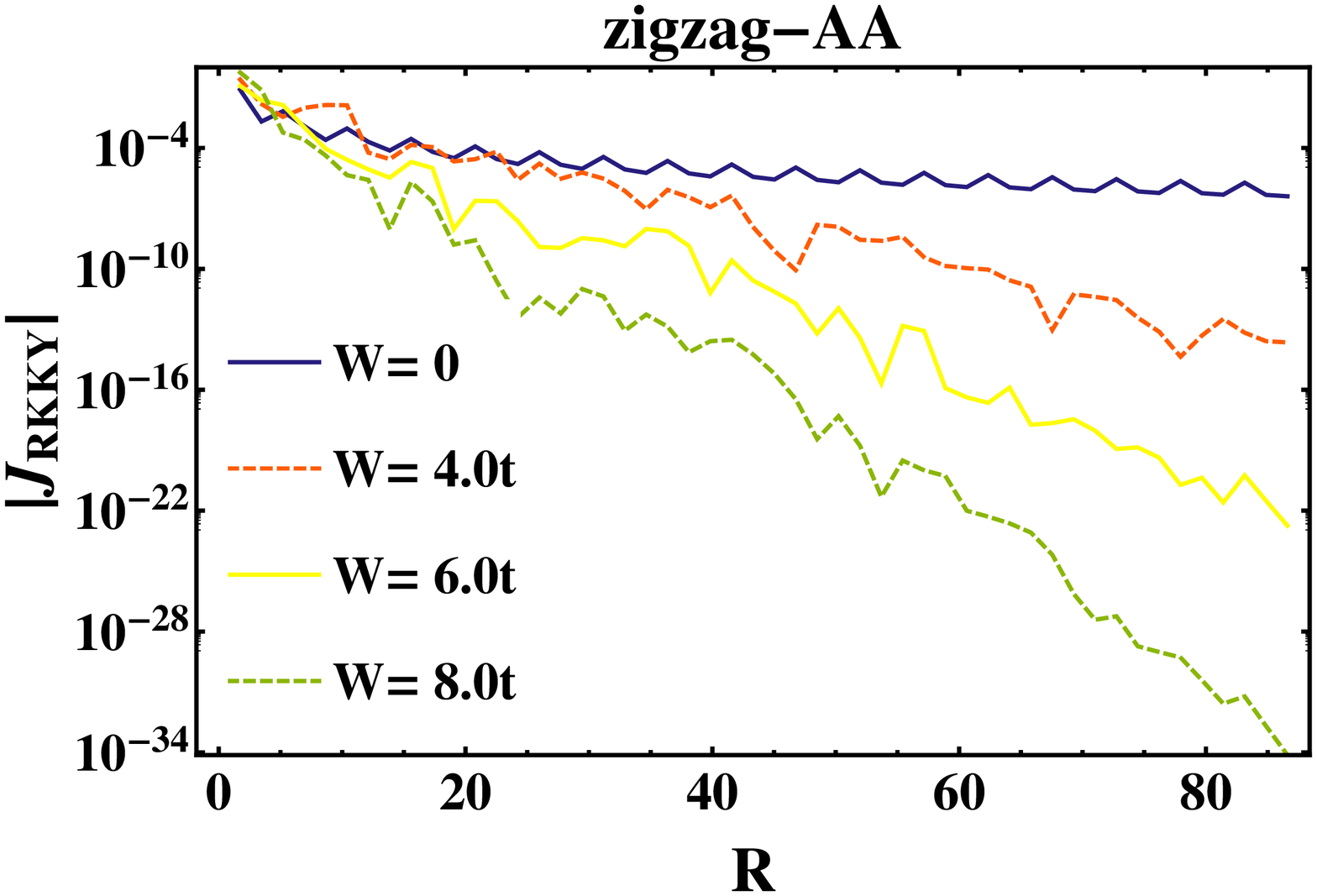}}\\
 \subfloat[~~~~~~~~~~~~~~~~~~~~~~~~~~~~~~~~~~~~~~~~~~~~~~~~~~~~] 
 {\includegraphics[width=0.4\textwidth]{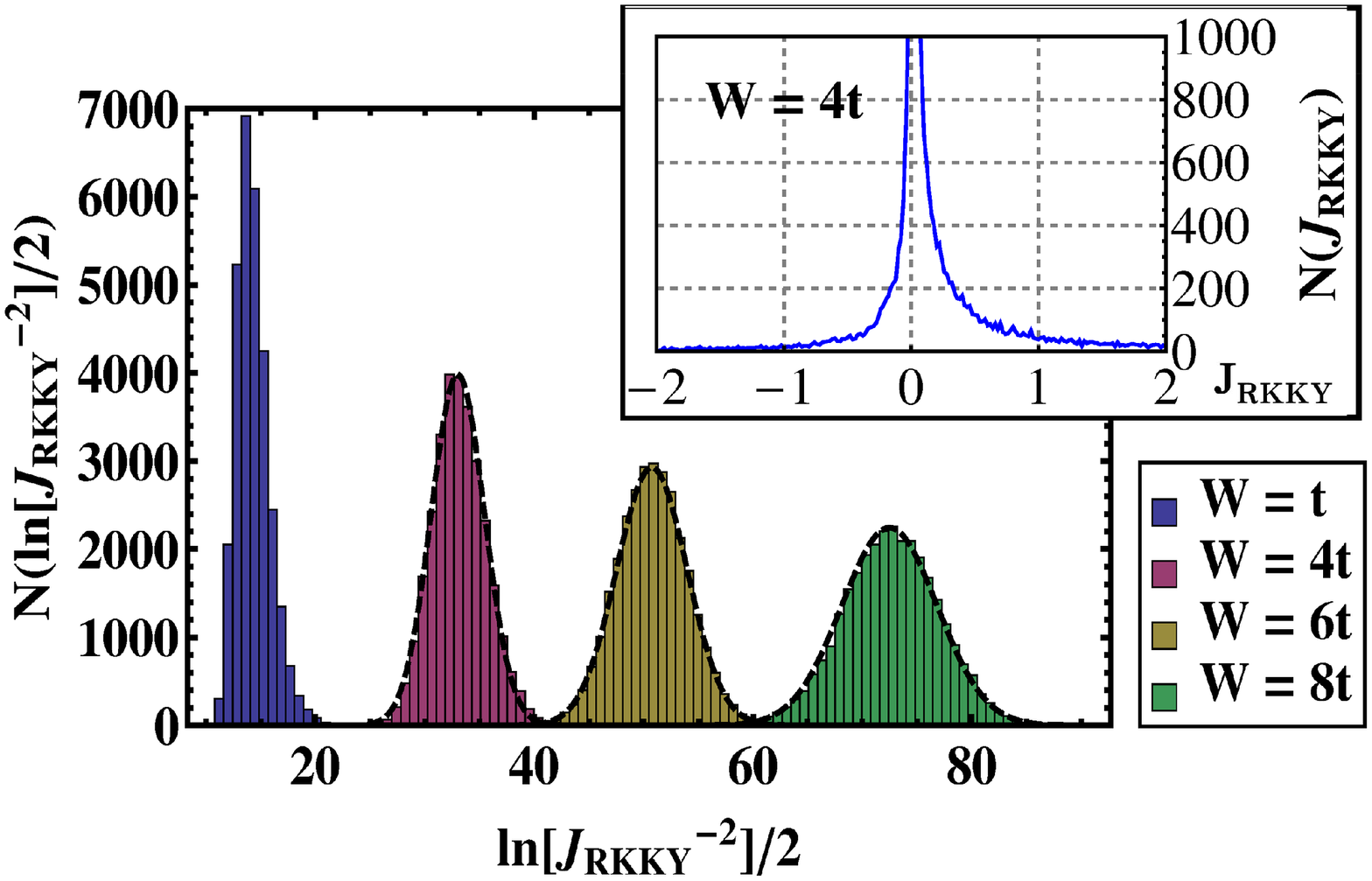}} 
 \caption{(Color online) Plots of (a) the absolute value of the RKKY
   interaction in a disordered sample as a function of distance and
   (b) of the distribution of the logarithm of the RKKY interaction
   amplitude for $R = 50\sqrt{3}a$ and different disorder
   strengths. The inset of (b) shows the distribution of the amplitude
   itself for $W=4t$. A lattice with $1.8\times 10^5$ sites and a
   polynomial degree cutoff of $M = 3\times 10^3$ are used. The black
   dashed lines represent the resulting distribution functions with
   $\sigma = 2.4,\,3.2,\,4.35$ and $\alpha = 32.5,\, 50,\, 72.5$ for
   $W = 4t,\, 6t,\, 8t$, respectively\,(see text).}
 \label{fig:distributions} 
\end{figure}

The typical value is found to have the same power-law decaying
behavior with distance as the amplitude of the RKKY interaction in the
clean limit.\cite{Zyuzin, Bergmann, Lerner, Sobota, Bulaevskii} Before
a direct evaluation of the geometrical average, we have calculated the
density of state for two disorder strengths \,$W = 0.5t, t$ to observe
how weak disorder affects the pseudogap at the neutrality point ($E =
0$), since, as we have seen above, the power of the pseudogap is
directly related to the unconventional power-law decay of the RKKY
interaction. For the density-of-states calculation, we used the KPM
\cite{Weisse, Amini} method with $3\times 10^7$ sites and a polynomial
degree cutoff of $M = 7\times 10^3$. As one may see from the plot of
the density of states in \,(Fig.\,\ref{fig:geo_avg}a), the pseudogap
is still not filled for weak disorder, has the same power law as in
the clean limit, and only the slope around the neutrality point is
changed,\cite{Amini}

\begin{equation}
  \rho(E) = \gamma |E|,
\end{equation}
where $\rho(E)$ is the density of states and $E$ is the energy
measured in units of the hopping amplitude $t$. The slope $\gamma$
depends on the disorder strength and is obtained by fitting the data
in Fig. \ref{fig:distributions}.

Using the Born and T-matrix approximations, an analytical study has
reported that there is a logarithmic correction to the density of
states around the neutrality point, yielding $\rho(E) = |E|\ln|E|$ in
the presence of disorder.\cite{HU} This logarithm correction does not
change the power of the distance dependence of the RKKY
interaction\,[Eq.\eqref{eq:charge_density}]. Consequently, the
geometrical average of the RKKY interaction is expected to have the
same power-law exponent as the clean system\,($1/R^3$). The direct
numerical calculation shown in Fig.\,\ref{fig:geo_avg}b strongly
supports this conclusion.


Figure\,\ref{fig:distributions}a shows how the RKKY interaction in a
strongly disordered sample gets suppressed by several orders of
magnitude when the strength of the nonmagnetic random potential is
increased. In order to investigate the broadening of the distribution,
we employed $3\times 10^4$ realizations of the disorder potential and
calculated the RKKY interaction amplitude for a fixed distance $R =
50\sqrt{3}$. The results are shown in
Fig.\,\ref{fig:distributions}b. The inset indicates that the
interaction strength follows a distribution with very long assymetric
tails. The squared amplitude\,($J_{\textrm{RKKY}}^2$) has a
distribution similar to log-normal with a width that increases with
disorder strength. Using a field-theoretical approach valid in the
metallic regime, Lerner found \cite{Lerner} that the increase in the
strength of the nonmagnetic disorder leads to a crossover in the shape
of the distribution function from a broad non-Gaussian with the very
long tails in the weak disorder regime to a completely log-normal
distribution in the region of strong disorder regime. This may explain
the crossover of the distribution which we observe from weak disorder
$W=t$ to strong disorder \,($W = 4t,\,6t,\,8t$). In order to directly
compare we have fitted the results with a log-normal functional form,
which is given by
\begin{equation}
  P(x) = \frac{N}{\sqrt{2\pi \sigma^2}}\,
  \exp\Big[-\frac{(x-\alpha)^2}{2\sigma^2}\Big],
  \label{eq:probability}
\end{equation}
where $N = 3\times 10^4$ is the number of realizations and $x =
\ln[J_{\tr{RKKY}}^{-2}]/2 $. The data are shown together with the
fitting curves in Fig.\,\ref{fig:distributions}b (black dashed lines).

\begin{figure}[!Ht]
  {\includegraphics[width=0.3\textwidth]{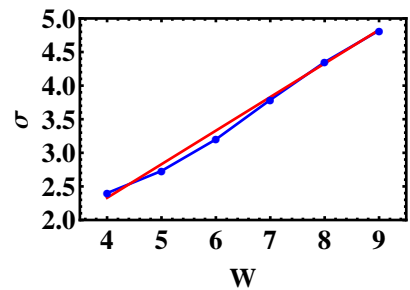}}
 \caption{(Color online) Plot of the width of the distribution of RKKY
   interaction amplitudes as a function of the disorder strength
   $W$\,(in unit of $t$). The red line represents a linear fitting
   curve\, which yields $\sigma = W/2 + 0.34$.}
  \label{fig:sigma} 
\end{figure}

The width\,$\sigma$ of the distribution in Eq. (\ref{eq:probability})
has been analyzed as functions of the disorder strength $W$ and is
shown in Fig.\,\ref{fig:sigma}. The red line in Fig.\,\ref{fig:sigma}
is a linear fitting curve ($\sigma = W/2 + 0.34$) and agrees with the
analytical prediction by Lerner, who has used the renormalization
group method to obtain \cite{Lerner}
\begin{equation}
  \sigma \sim 1/\sqrt{l_e} \sim W,
\end{equation}
where $l_e \sim 1/W^2$ is the mean free path of electrons, which we
studied in our previous work.\cite{HYLEE}

\section{Doped graphene\,($\mu\neq 0$)}
\subsection{Clean system\,($W=0$)}

\begin{figure}[!Ht]
 \captionsetup[subfloat]{font = {bf,up}, position = top,
   captionskip=0pt, farskip=0pt} 
 \subfloat[~~~~~~~~~~~~~~~~~~~~~~~~~~~~~~~~~~~~~~~~~~~~~] 
 {\includegraphics[width=0.35\textwidth]{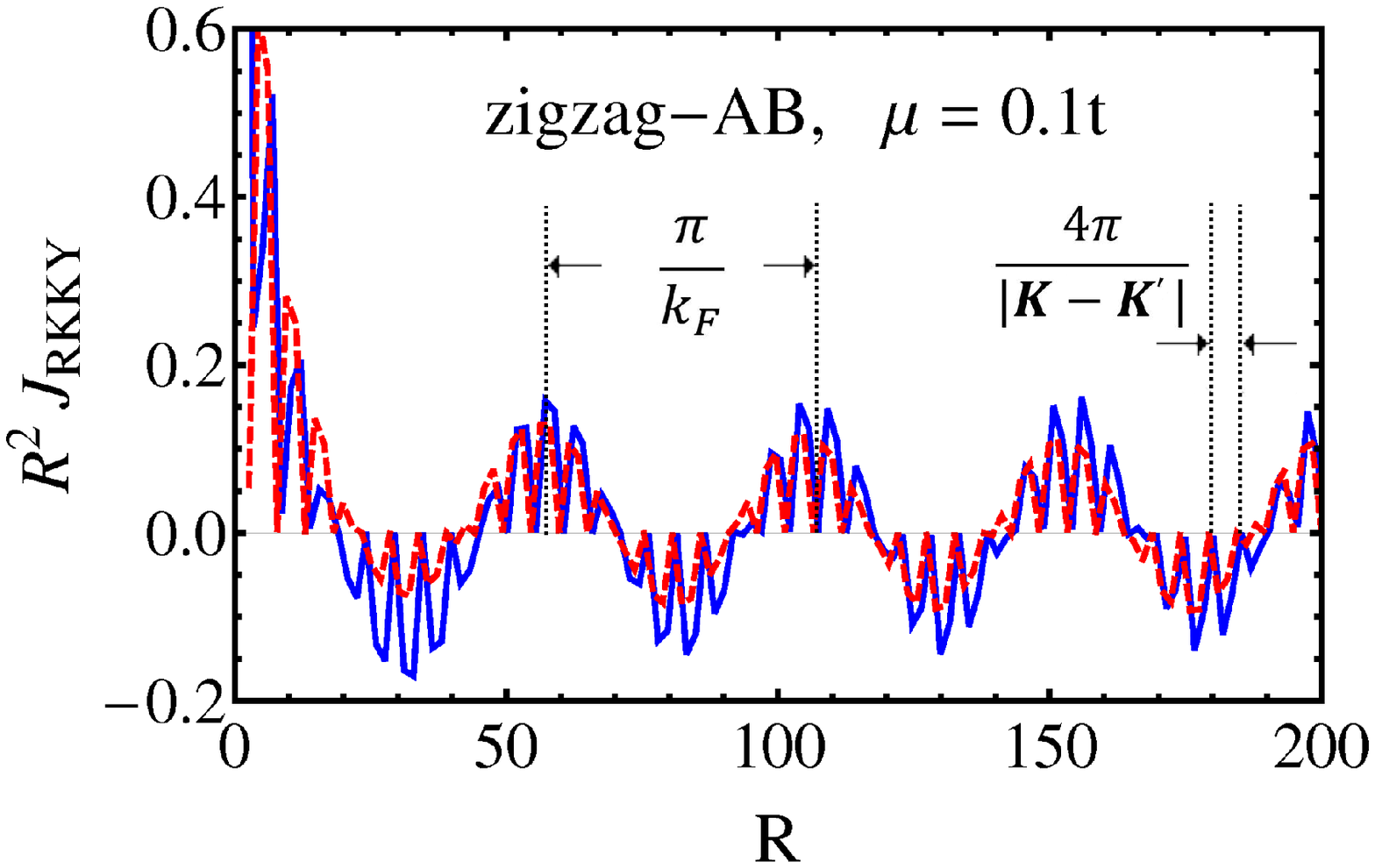}}\\
 \subfloat[~~~~~~~~~~~~~~~~~~~~~~~~~~~]
 {\includegraphics[width=0.24\textwidth]{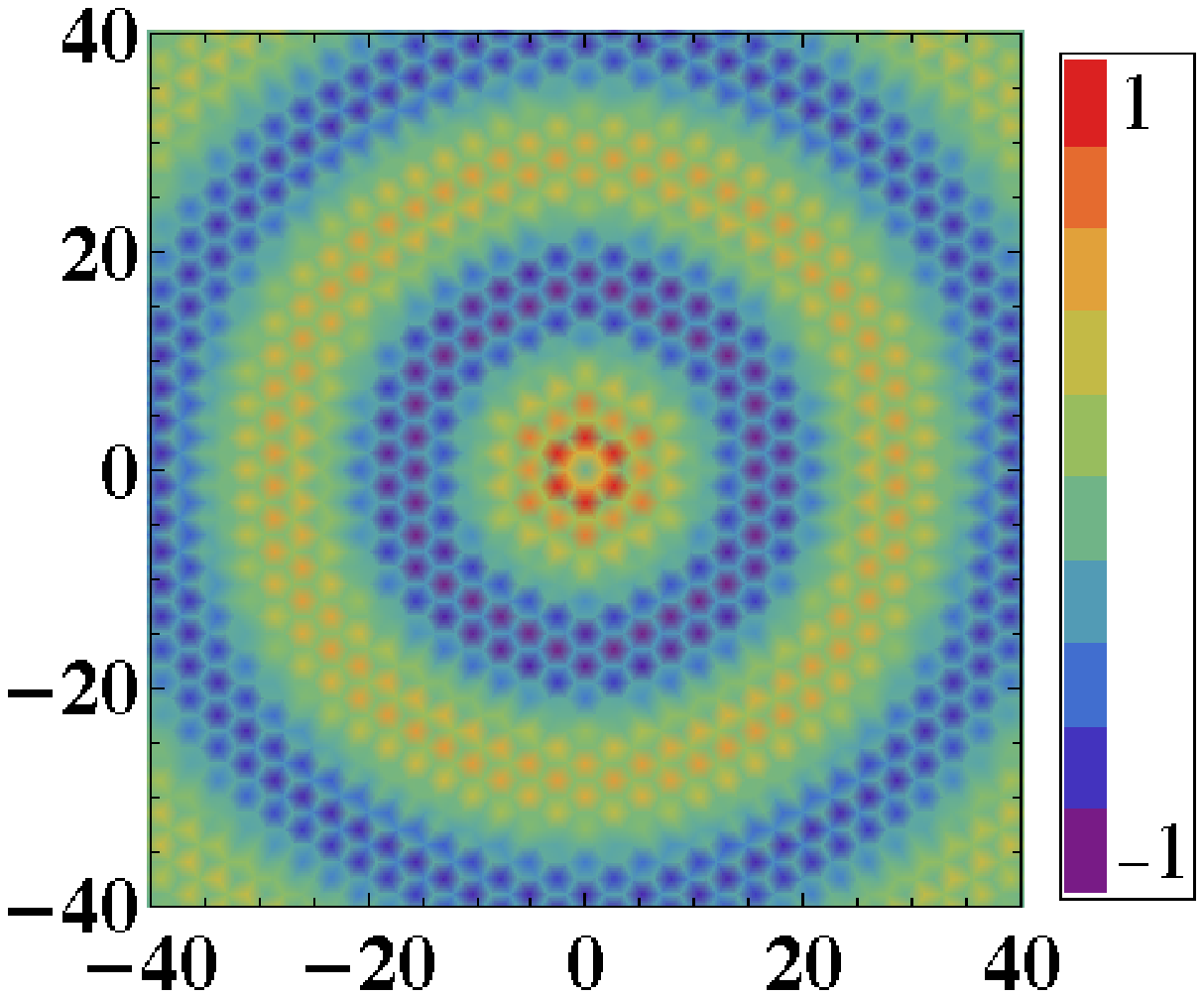}} 
 \subfloat[~~~~~~~~~~~~~~~~~~~~~~~~~~~]
{\includegraphics[width=0.24\textwidth]{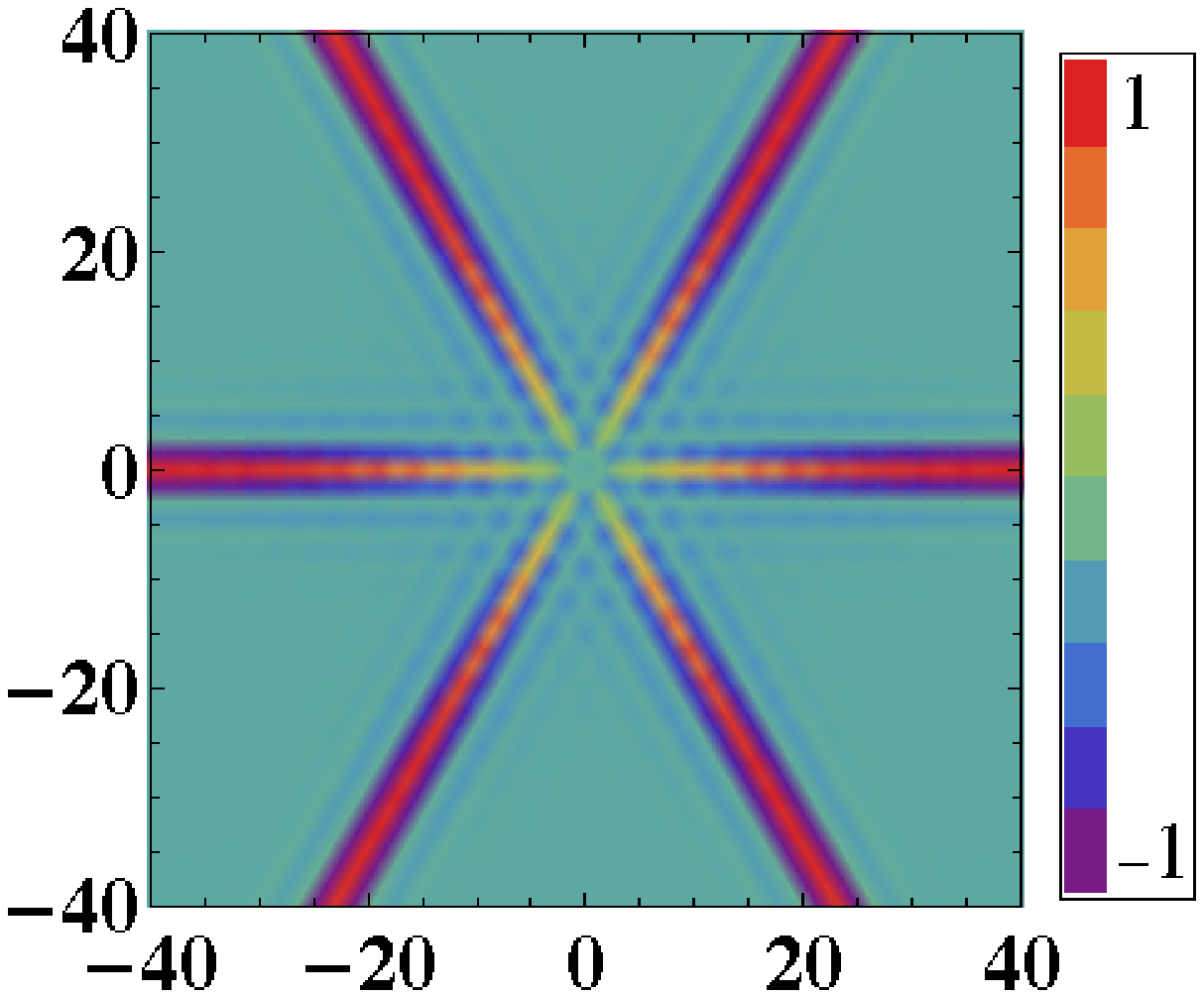}} 
 \caption{(Color online) (a) The RKKY interaction amplitude multiplied
   by $R^2$ in doped graphene at the chemical potential $\mu = 0.1t$
   along the zigzag-AB direction. Density plots of the RKKY
   interaction amplitude for (b) $\mu = 0.2t$ and (c) $\mu = t$ for
   all directions. In (a), the result of culations used the kernel
   polynomial method and the lattice Green's function method are
   represented as solid blue and dashed red line, respectively. A
   lattice with $7.2\times 10^5$ sites and a polynomial degree cutoff
   of $M = 5\times 10^3$ were used. The lattice constant $a$ is set to
   unity.}
 \label{fig:beatings} 
\end{figure}

By controlling $\mu$ in the Hamiltonian of
Eq.\,\eqref{eq:Hamiltonian}, we investigate how the RKKY interaction
evolves with the Fermi level. Results are shown in
Fig.\,\ref{fig:beatings}, where the interaction amplitude is
multiplied by $R^2$ in order to emphasize its oscillatory behavior. In
these calculations, we used a lattice with $7.2\times 10^5$ sites and
a polynomial degree cutoff $M = 3\times 10^3$. Near the neutrality
point a beating pattern appears as shown in Fig.\,\ref{fig:beatings}a,
b. It consists of a superposition of waves with the wave vectors
$\bm{K}-\bm{K}'$ and $\bm{q}_{F}$, where $\bm{q}_F$ is the Fermi wave
vector originating from the Friedel oscillations at finite Fermi
energy. Recently, the following analytical expressions for the beating
pattern were derived using lattice Green's functions:\cite{Sherafati1}
\begin{equation} 
  J_{\textrm{AA}} =  J_{\textrm{AA}}^{0} \bigg[ 1 +
    \frac{8 q_F R}
    {\sqrt{\pi}}G_{1,3}^{2,0}\left(\begin{matrix}\frac{1}{2},
        \frac{3}{2}\\ 1,1,1\end{matrix};q_{F}^{2}R^2
    \right)\bigg]
    \label{eq:rkky_near_diracAA}
\end{equation}
and
\begin{equation}
  J_{\textrm{AB}} = J_{\textrm{AB}}^{0} \bigg[ 1 - \frac{8q_F R} 
  {\sqrt{3\pi}}G_{2,4}^{2,1}\left(\begin{matrix}\frac{1}{2},  
      \frac{3}{2}\\ 1,2,0,-\frac{1}{2}\end{matrix};q_{F}^{2}R^2
  \right)\bigg], 
  \label{eq:rkky_near_dirac}
\end{equation}
where $J^{0}_{\textrm{AA(B)}}$ is the RKKY coupling function at the
neutrality point [see Eq. \eqref{eq:rkky_dirac}] and $G$ is the
Meijer-G function. Note that term within brackets describes the
isotropic dependence of the oscillations on the Fermi
momentum~$\bm{q}_F$. The external prefactor, $J^0_{\tr{AA(B)}}$, on
the contrary, is strongly anisotropic, depending on the vector given
by the momentum difference between the two neighbored Dirac points
$\bm{K}-\bm{K}'$. To make a comparison with our calculations, the
function represented by Eq.\,\eqref{eq:rkky_near_dirac} is also
presented in Fig. \ref{fig:beatings} by a red dashed line. Excellent
agreement is found. One can estimate the wavelength of the long
oscillation, which appears at finite Fermi level, using the dispersion
relation near the neutrality point, which is given by
\begin{equation}
  E(\bm{q}) = v_F | \bm{q} |,
  \label{eq:dispersion}
\end{equation}
where $\bm{q}$ is the momentum relative to the Dirac point and $v_F =
3ta/2$ is the Fermi velocity.\cite{Neto} The Fermi wavelength is found
to be about $\lambda_{F} \approx 50a$ for $\mu = 0.1t$. This coincides
with the period seen in \,(Fig.\,\ref{fig:beatings}a). As expected
from Eq.\,\eqref{eq:rkky_near_dirac}, the oscillations with large
period seen in (Fig.\,\ref{fig:beatings}b) are isotropic.

We have also calculated the RKKY interaction amplitude for highly
doped graphene\,($\mu = t$). The results are multiplied by the square
of the distance, $R^2$, and are shown in the density plots of
\,(Fig.\,\ref{fig:beatings}c). The behavior cannot be described by
Eqs.\,\eqref{eq:rkky_near_diracAA} and \eqref{eq:rkky_near_dirac}
which are valid only close to the neutrality point. When the Fermi
level is exactly at the van Hove singularity at \,($\mu = t$), the
ordering pattern of the RKKY interactions along the zigzag direction
is reversed. In other words, the correlation between impurities on
zigzag-AA or BB is always antiferromagnetic and ferromagnetic for
zigzag-AB or BA pairs. At the same time, the interactions are strongly
suppressed for the other directions. This is in accordance with a
result of a previous study.\cite{Bruno}

\subsection{Disordered system\,($W\neq 0$)}

\begin{figure}[!Ht]
 \captionsetup[subfloat]{font = {bf,up}, position = top,
   captionskip=0pt, farskip=-3pt} 
 \subfloat[~~~~~~~~~~~~~~~~~~~~~~~~~~~~~~~~] 
 {\includegraphics[width=0.24\textwidth]{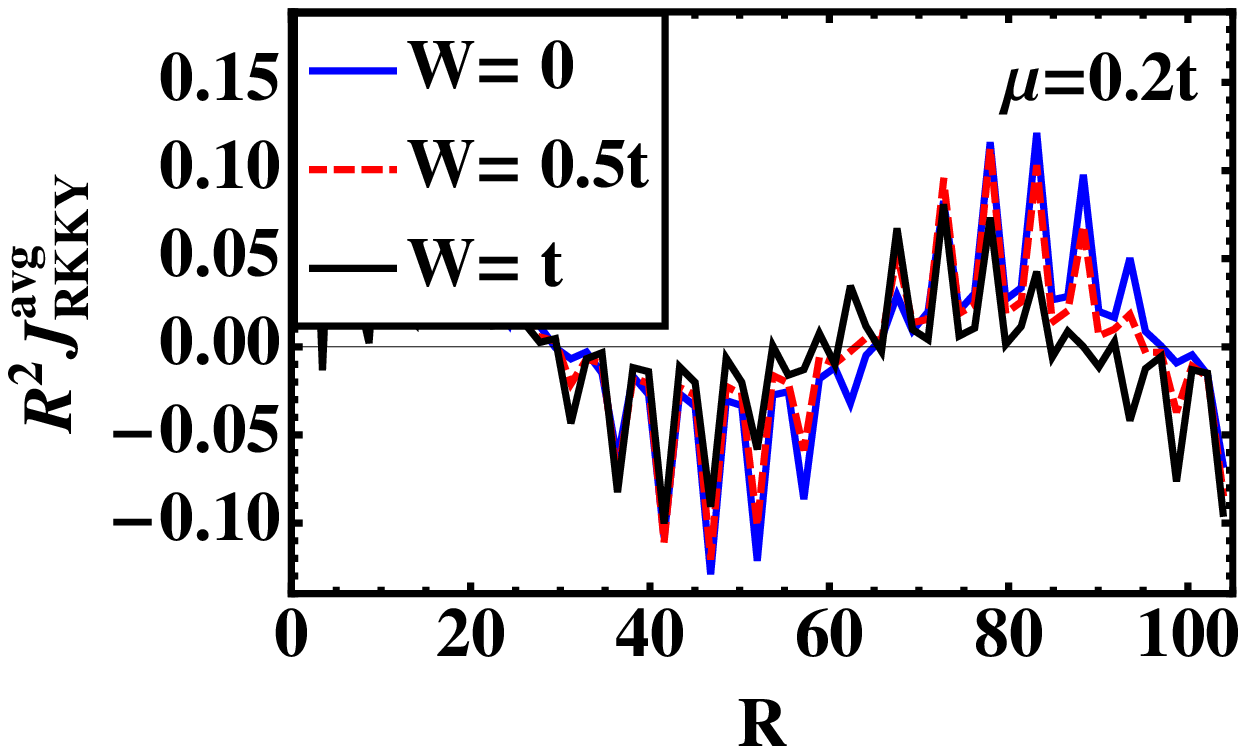}}
 \subfloat[~~~~~~~~~~~~~~~~~~~~~~~~~~~~~~~~] 
 {\includegraphics[width=0.24\textwidth]{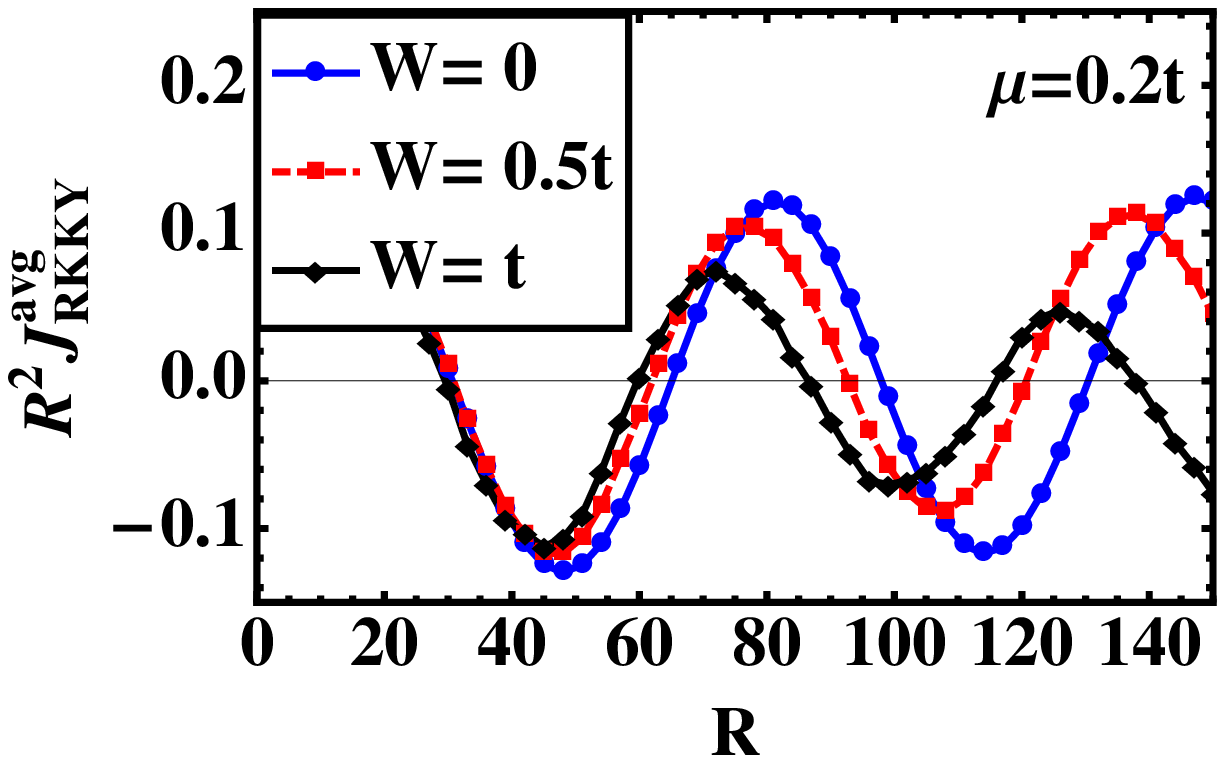}}
 \caption{(Color online) The average RKKY interaction strength
   multiplied by the square of the distance, $R^2$, along the (a)
   zigzag-AA and (b) armchair-AA directions at the Fermi energy
   $\mu=0.2t$ is plotted in the diffusive regime. 1600 different
   disorder configurations are used in the averaging. The same lattice
   size and polynomial cutoff as in Fig.\,\ref{fig:avg_angle} were
   used.}
 \label{fig:avg_mu} 
\end{figure}

We have also calculated the RKKY interaction amplitude in disordered
doped graphene and the results are shown in Fig.\,\ref{fig:avg_mu}. A
lattice with $5\times 10^5$ sites and a polynomial degree cutoff of $M
= 5\times 10^3$ were used in these numerical calculations. In our
previous work,\cite{HYLEE} we reported that the amplitude of the
averaged RKKY interaction along the zigzag-AA direction in the
ballistic regime\,($R<l_e$) increases with weak
disorder\,(Fig.\,\ref{fig:distributions_weak}f). Even though the
unusual oscillation coming from the interference of two Dirac points
still exists\,(Fig.\,\ref{fig:avg_mu}a), the amplitude of the envelope
of the averaged RKKY interaction decreases with disorder strength\,$W$
and the period\,($2\pi/k_{\tr{F}}$) of the oscillation coming from the
finite Fermi energy is modified by the disorder as in a conventional
metal.

\section{Discussion and Conclusions}

In conclusion, we have employed a semiclassical description of the
RKKY interaction in terms of the modulation of the charge density by
the presence of magnetic impurities in order to rederive an analytical
expression applicable to pure graphene. This semiclassical approach
provides not only a simpler derivation, but also an intuitive
interpretation of the origin of the oscillating behavior of the RKKY
interaction at zero doping as an interference of the two degenerate
Dirac points. Moreover, the origin of the unusual power-law decay of
the RKKY interaction in pure graphene at the neutrality point is
clearly related to the pseudogap in the density of states. Using a
Feynman's path integral scheme, we could also trace the origin of the
anisotropic sensitivity of the RKKY interaction to disorder to the
presence of multiple shortest paths between two magnetic impurities in
the armchair direction, showing that this direction is more robust to
disorder than the zigzag one.

As an extension of our previous study,\cite{HYLEE} we have calculated
and studied the RKKY interaction in doped and disordered graphene in
detail using the kernel polynomial method. Fnite gate voltage breaks
particle-hole symmetry and the resulting finite Fermi surface yields
Friedel oscillations, so that the sign of the RKKY interaction between
the impurities localized on the same sublattice now oscillates with
distance. When the Fermi level is exactly at the van Hove
singularity\,($\mu = t$), the ordering pattern of the RKKY
interactions along the zigzag direction is reversed. At the same time,
the interactions are strongly suppressed for all other directions.

In order to study the anisotropic influence of nonmagnetic disorder, we
evaluated the RKKY interactions along two different directions between
the zigzag and armchair directions. As reported in our previous study
and expected from the semiclassical approach, in the diffusive regime
the armchair direction is not affected by the nonmagnetic disorder as
much as the other directions. We have also found that, in the
ballistic regime ($R<l_e$), the distribution of the RKKY interactions
along the zigzag direction is not universal but depends on the lattice
sites at which the pair of magnetic impurities sit. We identified
three different representative shapes, which repeat themselves
periodically. By an accurate evaluation of the density of states
around the neutrality point in weakly disordered regime\,($W\leq t$),
we confirmed that the linear dispersion relation is still valid and the
pseudogap is not filled. This is in full agreement with the fact that
the geometrical average of the RKKY interaction in the diffusive
regime decays as in the clean system, namely, as $1/R^3$, and not
as $1/R^2$, which is the usual behavior for two-dimensional metals. In
the localized regime\,($R>\xi$), the geometrical average is also
exponentially suppressed at distances exceeding the localization
length $\xi$ and the distribution of the strength of the RKKY
interaction shows a crossover from the non-Gaussian shape with very
long tails to the completely log-normal form when increasing the
disorder strength. We have analyzed the width of the log-normal
distribution and confirmed that it increases linearly with the
amplitude of the disorder potential $W$.

\begin{figure}[!Ht]
\captionsetup[subfloat]{font = {bf,up}, position = top,
 captionskip=0pt, farskip=0pt} 
\subfloat[~~~~~~~~~~~~~~~~~~~~~~~~~~~~~~~~~~~~~~~~~~~~~~~~~~~~~~~~] 
{\includegraphics[width=0.38\textwidth]{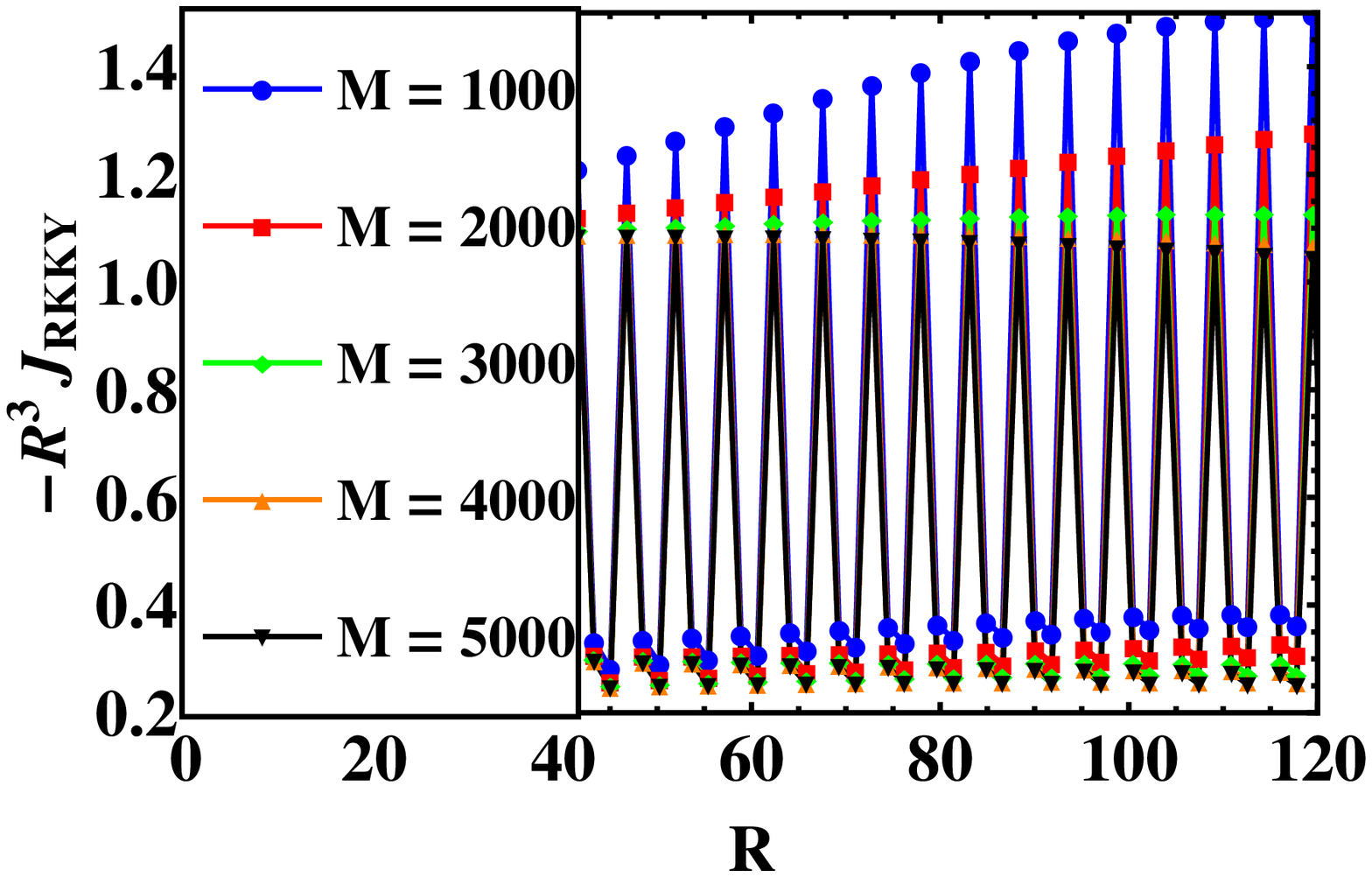}}\\
\subfloat[~~~~~~~~~~~~~~~~~~~~~~~~~~~~~~~~~~~~~~~~~~~~] 
{\includegraphics[width=0.3\textwidth]{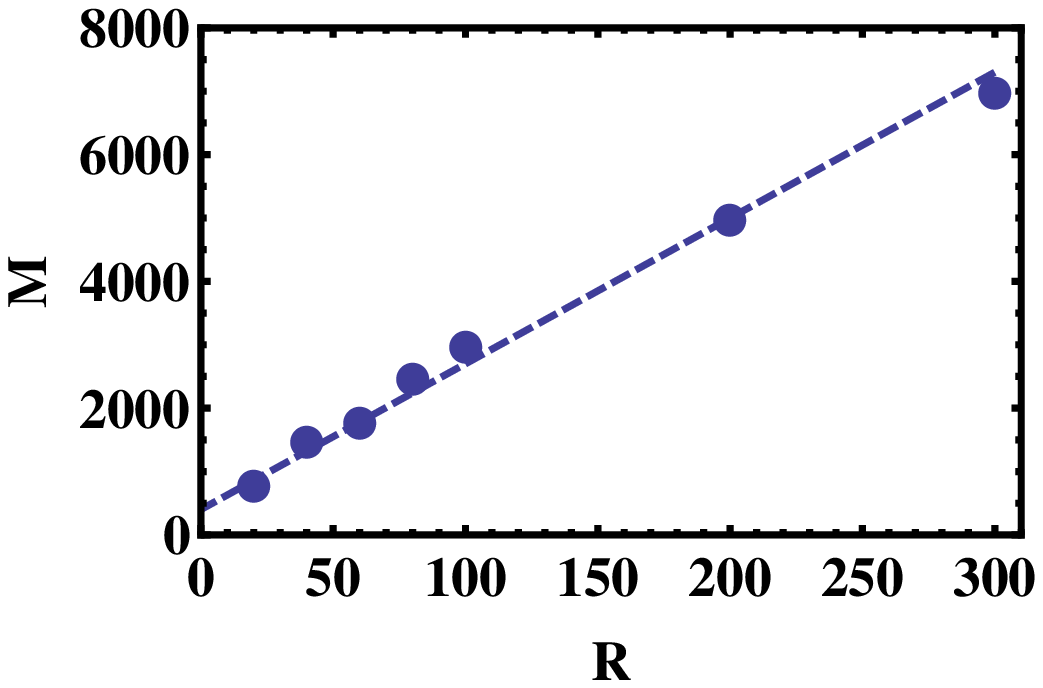}} 
\caption{(Color online) Plots of (a) the RKKY interaction as function
  of the polynomial degree $M$ in a lattice with $5\times 10^5$ sites.
  (b) The smallest cutoff number $M$ that yields a convergent result
  as function of the distance between the magnetic impurities $R$.}
\label{fig:kpm1} 
\end{figure}

In this work we showed that the KPM method is efficient and accurate
for studying interactions in disordered systems. In order to minimize
the computation time while keeping the highest accuracy, the
convergence of the calculations with respect to the Chebyshev
polynomial cutoff degree $M$ and the system size $L$ have been
investigated in detail (see Appendix A). The proper cutoff $M$ to
reach convergence is found to increase linearly with the distance $R$
between two magnetic impurities, as seen in Fig. \ref{fig:kpm1} b. For
a given distance $R$, a system size $L$ about five times larger than
$R$ is found to give convergent results. In comparison with the exact
numerical diagonalization, the KPM is found to be much faster. It can
be implemented for very large system sizes, even in disordered ones,
where thousands of realizations are needed in order to yield
meaningfull statistics.

\appendix

\section{Convergence of the Kernel Polynomial Method Calculations}

When using the KPM, the calculation of the Chebyshev polynomials using
recurrence relations consumes most of the computation time. Therefore,
we investigated first the relation between the cutoff number\,$M$ and
the convergence of the results in order to be able to minimize $M$ and
optimize the calculations. For clean graphene, a lattice with $5\times
10^5$ sites was used in these calculations. When a cutoff number $M$
is not sufficient, the amplitude of the RKKY interaction deviates from
the expected power-law behavior, as indicated by the blue and red
lines in Fig.\,\ref{fig:kpm1}a). When we determined the smallest
cutoff number $M$ such that the variance of the amplitude of the RKKY
interaction is less than 5\%, we found that it increases linearly with
the distance between two magnetic impurities $R$ as seen in
Fig.\,\ref{fig:kpm1}b). This linear relation between the distance and
the cutoff number allows the rapid calculation of the RKKY interaction
amplitude between the magnetic moments even at large distances $R$.

\begin{figure}[!Ht]
  \captionsetup[subfloat]{font = {bf,up}, position = top,
    captionskip=0pt, farskip=-5pt} 
  \subfloat
  {\includegraphics[width=0.3\textwidth]{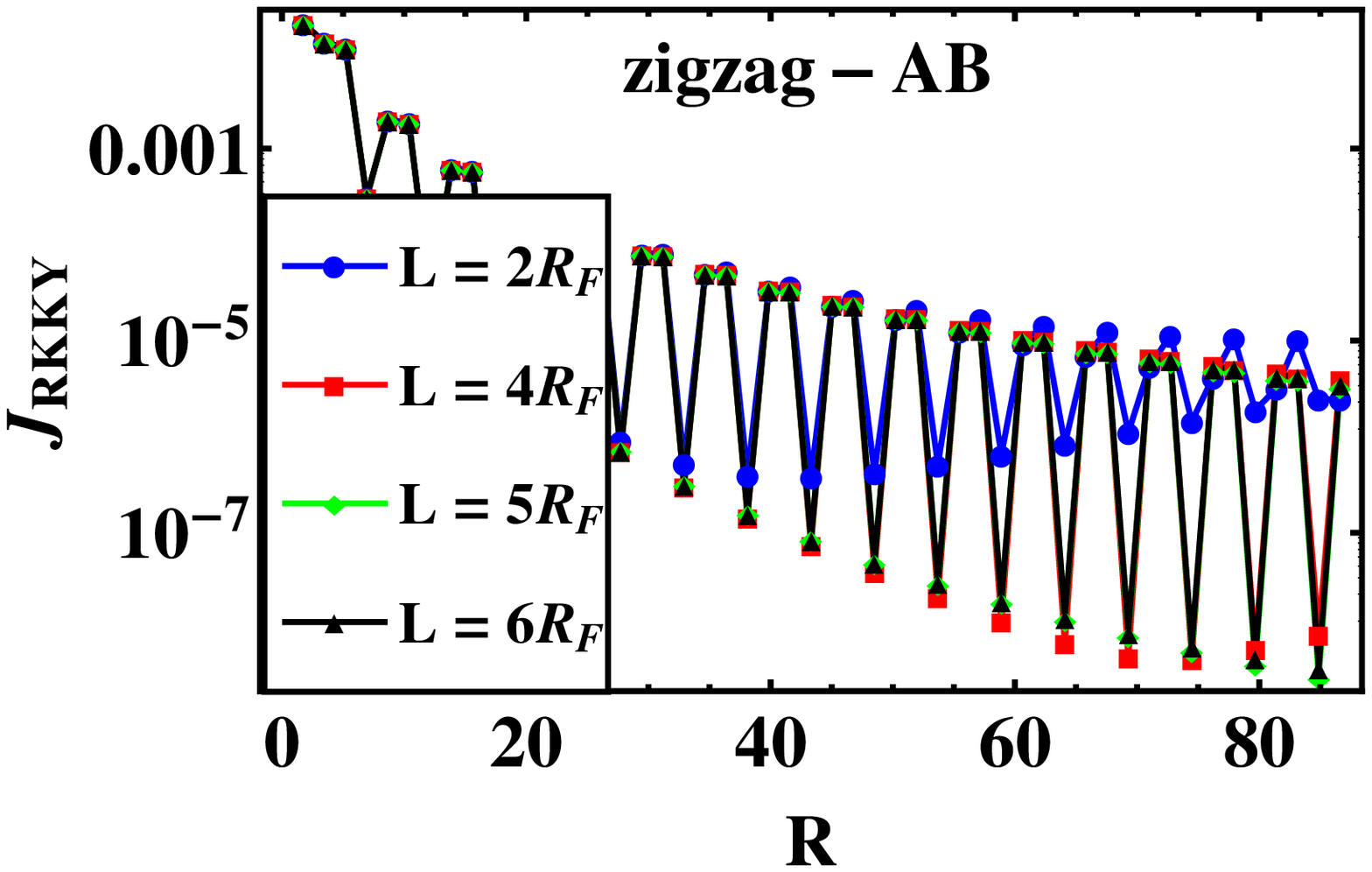}}
  \subfloat
  {\includegraphics[width=0.2\textwidth]{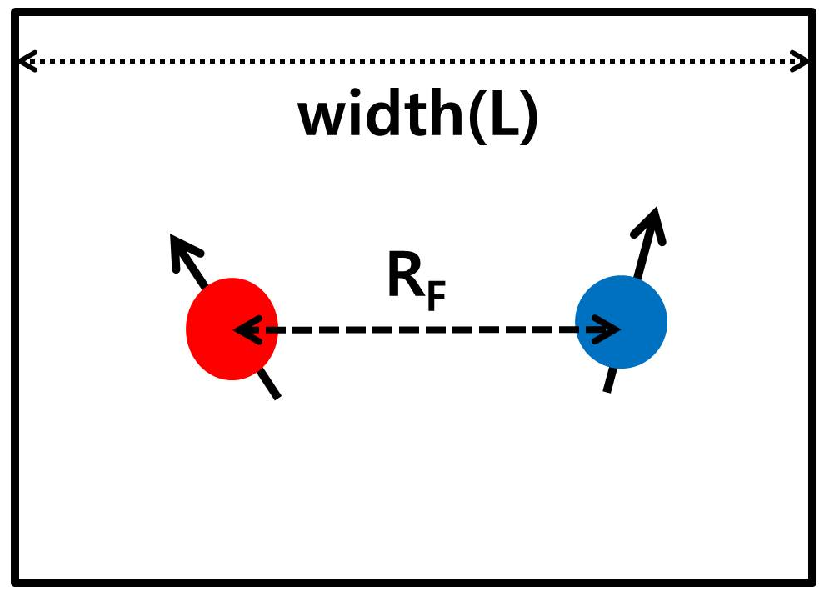}} 
  \caption{(Color online) Plots of the RKKY interaction in terms of a
    system size\,$L$. $50\sqrt{3} a$ is used as the longest distance
    $R_F$. A cutoff number $M=5\times 10^3$ is used in these
    calculations.}
  \label{fig:kpm2} 
\end{figure}

In order to minimize the KPM calculation time further, we have also
studied the smallest system size $L$ which yields convergent results,
as shown in Fig.\,\ref{fig:kpm2}. $R_F$ denotes the longest distance
used in the calculation, $R_F = 50\sqrt{3}a$ and the cutoff number is
$M=5\times 10^3$. One can observe a good convergent behavior\,(see
green and black line in Fig.\,\ref{fig:kpm2}) when the system size $L$
is larger than $5R_F$. The exact diagonalization method also yields a
proper result when $L=5R_F$.\cite{Annica} However, the KPM does not
require matrix diagonalization and therefore is a much faster
computational tool.

\acknowledgments

This research was supported by WCU (World Class University) program
through the National Research Foundation of Korea funded by the
Ministry of Education, Science and Technology(Grant
No. R31-2008-000-10059-0), Division of Advanced Materials
Science. E.R.M. acknowledges partial support through the NSF DMR
(Grant No. 1006230). E.R.M. and G.B. thank the WCU AMS for its
hospitality. H. Y. Lee thanks Jacobs University Bremen for its
hospitality.

\bibliographystyle{apsrev}

\bibliography{reference}

\end{document}